\def\year{2023}\relax
\documentclass[letterpaper]{article} 
\usepackage{aaai23}  
\usepackage{times}  
\usepackage{helvet}  
\usepackage{courier}  
\usepackage[hyphens]{url}  
\usepackage{graphicx} 
\usepackage{natbib}  
\usepackage{caption} 
\DeclareCaptionStyle{ruled}{labelfont=normalfont,labelsep=colon,strut=off} 
\usepackage{algorithm}
\usepackage{algorithmic}

\usepackage{xcolor}

\usepackage{newfloat}
\usepackage{listings}
\floatstyle{ruled}
\newfloat{listing}{tb}{lst}{}
\floatname{listing}{Listing}
\nocopyright
\usepackage{makecell}
\usepackage{bm}
\usepackage{enumitem}
\usepackage{multirow}
\usepackage{tabularx}
\usepackage{amssymb}
\usepackage{amsmath}
\usepackage{pifont}
\usepackage{booktabs}
\usepackage{graphicx}
\usepackage{nameref}
\usepackage{lineno}
\nolinenumbers
\usepackage[subrefformat=parens,skip=0pt]{subcaption}
\usepackage[detect-all]{siunitx}
\usepackage{xspace}
\newcommand\ie{i.\,e.\xspace}
\newcommand\eg{e.\,g.\xspace}

\def\sym#1{\ifmmode^{#1}\else\(^{#1}\)\fi}

\let\namerefOld\nameref
\renewcommand{\nameref}[1]{\textit{\namerefOld{#1}}}

\newsavebox{\measurebox}

\title{Community Fact-Checks Do Not Break Follower Loyalty}

\author{
    Michelle Bobek\textsuperscript{\rm 1}, Nicolas Pr{\"o}llochs\textsuperscript{\rm 1}
}
\affiliations{
    \textsuperscript{\rm 1}JLU Giessen, Germany \\
    michelle.j.bobek@uni-giessen.de, nicolas.proellochs@wi.jlug.de
}

\begin{document}

\maketitle

\begin{abstract}
Major social media platforms increasingly adopt community-based fact-checking to address misinformation on their platforms. While previous research has largely focused on its effect on engagement (\eg, reposts, likes), an understanding of how fact-checking affects a user's follower base is missing. In this study, we employ quasi-experimental methods to \emph{causally} assess whether users lose followers after their posts are corrected via community fact-checks. Based on time-series data on follower counts for $N=3516$ community fact-checked posts from X, we find that community fact-checks do \emph{not} lead to meaningful declines in the follower counts of users who post misleading content. This suggests that followers of spreaders of misleading posts tend to remain loyal and do not view community fact-checks as a sufficient reason to disengage. Our findings underscore the need for complementary interventions to more effectively disincentivize the production of misinformation on social media.
\end{abstract}

\section{Introduction}

The spread of online misinformation has become a defining challenge of the digital age \cite{WEF.2024}. Misleading claims have been repeatedly linked to detrimental outcomes in various domains, including elections \cite{Allcott.2017, Bakshy.2011, McCabe.2024}, public health \cite{Gallotti.2020, Pennycook.2020b, Solovev.2022b}, and public safety \cite{Baer.2022,Starbird.2014, Oh.2013}. Social media platforms, by virtue of their design and scale, have become fertile ground for the dissemination of such content \cite{Bar.2023}. Consequently, the development of effective countermeasures to mitigate the spread of online misinformation has become an urgent task. 

For years, social media platforms have relied on professional third-party fact-checking organizations, such as \emph{snopes.com} or \emph{politifact.com}, where expert reviewers assess the accuracy of online claims and publish corrections \cite{Wu.2019, Vosoughi.2018, Pilarski.2024}. While such expert-led efforts tend to be highly accurate, they are often criticized for being too slow and limited in coverage to keep pace with the speed and scale of misinformation online. Furthermore, many users perceive professional fact-checks as politically biased, which has led to growing distrust in expert fact-checking \cite{Poynter.2019, Drolsbach.2024}. 

To address these challenges, social media platforms have begun to explore community-based fact-checking as an alternative. Unlike expert-led efforts, community-driven approaches leverage the collective judgment of platform users, \ie non-experts, to identify and correct misleading content \cite{Allen.2021, Bhuiyan.2020, Pennycook.2019, Prollochs.2022a, Drolsbach.2023, Drolsbach.2023b}. This strategy builds on the principle known as the ``wisdom of crowds'' -- the idea that the aggregated assessments of non-experts can approximate expert-level accuracy \cite{Frey.2021}. Building on this principle, X (formerly Twitter) has introduced Community Notes, a crowd-sourced fact-checking feature that allows users to annotate potentially misleading posts \cite{Twitter.2021,Prollochs.2022a}. A note only gets displayed underneath the original post, when it receives helpful ratings from users with diverse perspectives, to mitigate the risk of political or ideological viewpoints from dominating \cite{Solovev.2025}. Prior studies indicate that harnessing crowd intelligence can increase the speed and volume of fact-checking \cite{Pennycook.2019, Chuai.2024b}, and that users perceive community notes as more trustworthy than traditional fact-checks \cite{Drolsbach.2024}. 

While prior work has demonstrated that community fact-checks can generate high-quality fact-checks at scale, research on how users respond to these corrections is still in its infancy. The few existing studies in this direction have mainly focused on post-level engagement, finding that community notes reduce likes, reposts, and replies to flagged posts \cite{Chuai.2024b, Slaughter2025}. However, little is known about the potential reputational consequences for the authors of the fact-checked posts. In particular, it remains unclear whether being corrected via community notes affects a user's follower base -- for example, by prompting others to unfollow or disengage. Understanding this is important because reputational costs may function as a key behavioral incentive: if being publicly corrected has no consequences on social ties, the ability of community-based fact-checking to discourage misinformation may be limited.

From both a theoretical and empirical perspective, the effect of community notes on follower counts is unclear, with plausible arguments for both follower loss and retention. On the one hand, sharing false information can cause reputational harm \cite{Altay.2020b} and being fact-checked may damage a user’s perceived credibility. This may lead some followers (particularly those who value accuracy) to disengage. On the other hand, however, many users follow accounts based on social factors and ideological alignment \cite{Aiello.2012, Barbera.2015a} rather than factual accuracy \cite{Ashkinaze.2024}. This suggests that such audiences may be less responsive to public corrections, making it unlikely that being community fact-checked results in substantial follower loss. Existing empirical evidence outside of the context of community-based fact-checking reflects this tension. Here, surveys suggest that users intend to unfollow peers who spread misinformation, particularly when it clashes with their political views \cite{Kaiser.2022}. Yet observational research on expert-based fact-checking shows that misinformation spreaders are less likely to be unfollowed than non-spreaders \cite{Ashkinaze.2024}. These mixed results highlight the absence of a consistent understanding of the user-level consequences of being publicly fact-checked -- even outside the specific context of community-based fact-checking -- and the need for causal evidence on how fact-checking affects follower dynamics.

\textbf{Research goal:} In this study, we \emph{causally} analyze whether authors of misinformation lose followers once their posts are corrected via community fact-checks. To this end, we compile a unique longitudinal dataset comprising $N=3516$ posts that have been fact-checked via X's Community Notes platform between March and September 2024, \ie, within an observation period of seven months.  For each post, we track daily follower counts over a 21-day window centered around the post’s publication.  To estimate causal effects, we leverage variation in the timing of community notes and apply a staggered difference-in-differences (DiD) design. This enables us to isolate the causal effect of community notes on daily follower counts. 

\textbf{Contribution:} To the best of our knowledge, our study is the first to causally analyze whether authors of misinformation lose followers after their their posts are corrected via community notes. Across a wide range of model specifications, outcome measures, and sub-samples, we find that community fact-checks do \emph{not} lead to meaningful declines in follower counts. This indicates that such corrections may have limited influence on social ties and highlights the need for complementary strategies to more effectively disincentivize the production of misinformation on social media.

\section{Background}

\subsection{Misinformation on social media}

With more than half of U.S. adults consuming news via these platforms, social media has become a central outlet for information dissemination and public discourse \cite{VanBavel.2024, Pew.2024b}. Their growing appeal stems from convenience, speed, and the social nature of news sharing \cite{Pew.2024a}. Given that anyone can post and share content, social media facilitates rapid and large-scale diffusion of information \cite{Lazer.2018, Shore.2018, Kim.2019}. However, unlike traditional media, social media lacks oversight by experts, leaving little control over the content spreading. This renders these platforms particularly vulnerable to the spread of misinformation \cite{Shao.2016, Vosoughi.2018}. The findings of several studies on the diffusion of misinformation suggest that it can spread more viral than the truth \cite{Vosoughi.2018, Solovev.2022b, Prollochs.2021a, Prollochs.2022b}. Exposure to misinformation on social media has been associated with adverse outcomes, including misperceptions during elections \cite{Allcott.2017, Bakshy.2015, McCabe.2024}, and risky behaviors during public health crises \cite{Gallotti.2020, Pennycook.2020b, Solovev.2022b}. This challenge is further exacerbated by advances in AI enabling the generation of misinformation at increasing speed and scale \cite{Feuerriegel.2023}. 

Existing research highlights several factors contributing to the spread of misinformation online. For instance, misinformation is often written to intentionally mislead, complicating users' ability to recognize false information \cite{Wu.2019}. Moreover, social media users rarely verify the accuracy of the content they encounter \cite{Geeng.2020, Vo.2018}, suggesting that many lack the cognitive reflection needed to critically assess content accuracy \cite{Moravec.2019, Pennycook.2019b, Pennycook.2021}. Additionally, online social networks tend to reflect homophily, the tendency for individuals to associate with others sharing similar beliefs \cite{McPherson.2001}. This dynamic facilitates the formation of echo chambers, in which users are predominantly exposed to like-minded perspectives \cite{Barbera.2015}. Within such environments, misinformation is less likely to be challenged and may be reinforced through repeated exposure, further strengthening false beliefs \cite{Pennycook.2018}.

\subsection{Community-based fact-checking}

\begin{figure*}[t]
\centering
\sbox{\measurebox}{
  \begin{minipage}[b]{.47\textwidth}
    \centering
    \includegraphics[width=\textwidth]{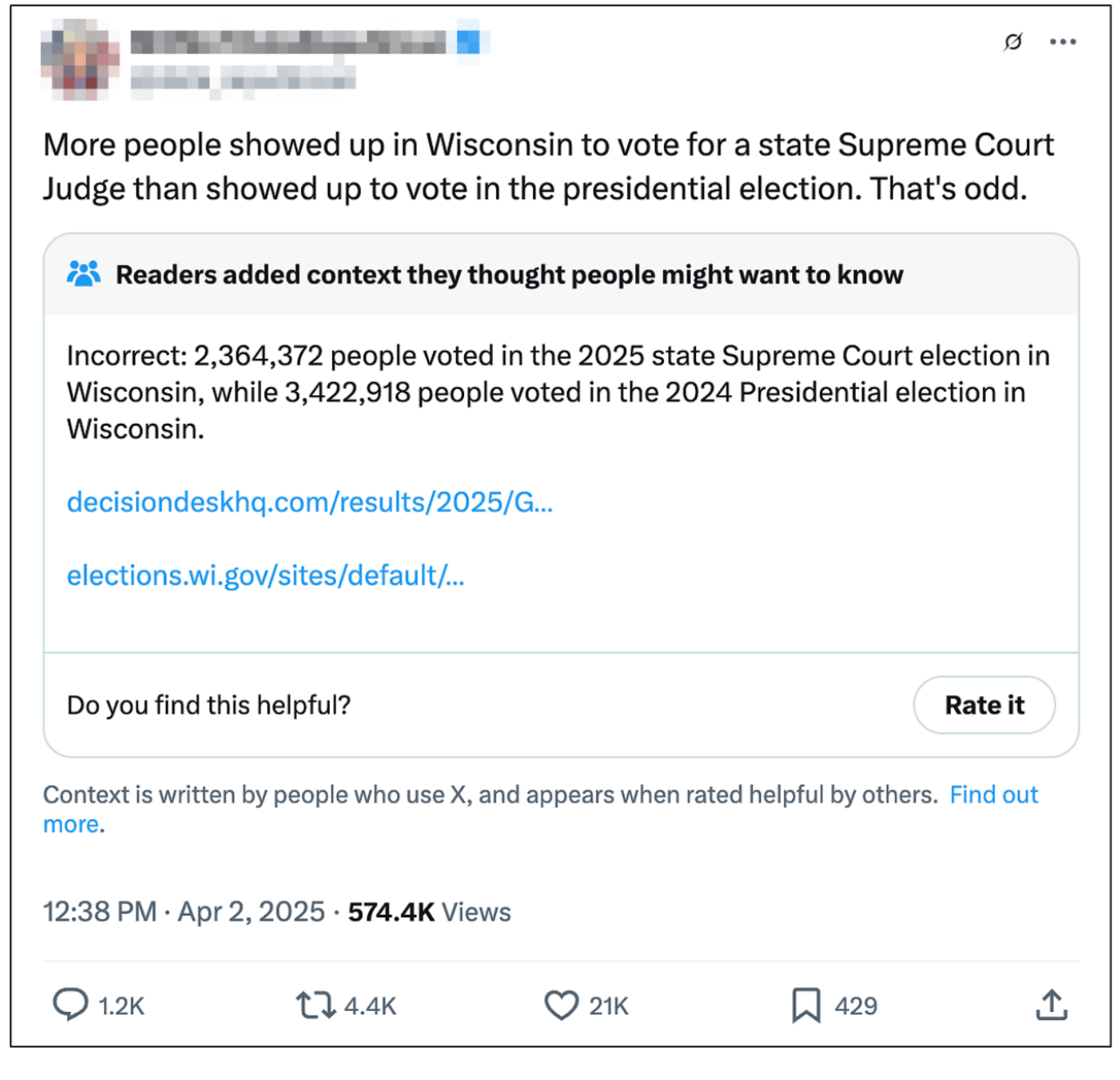}
    \subcaption{} 
    \label{fig:note_example}
  \end{minipage}}
\usebox{\measurebox}\qquad
\begin{minipage}[b][\ht\measurebox][s]{.47\textwidth}
    \centering
    \includegraphics[width=\textwidth]{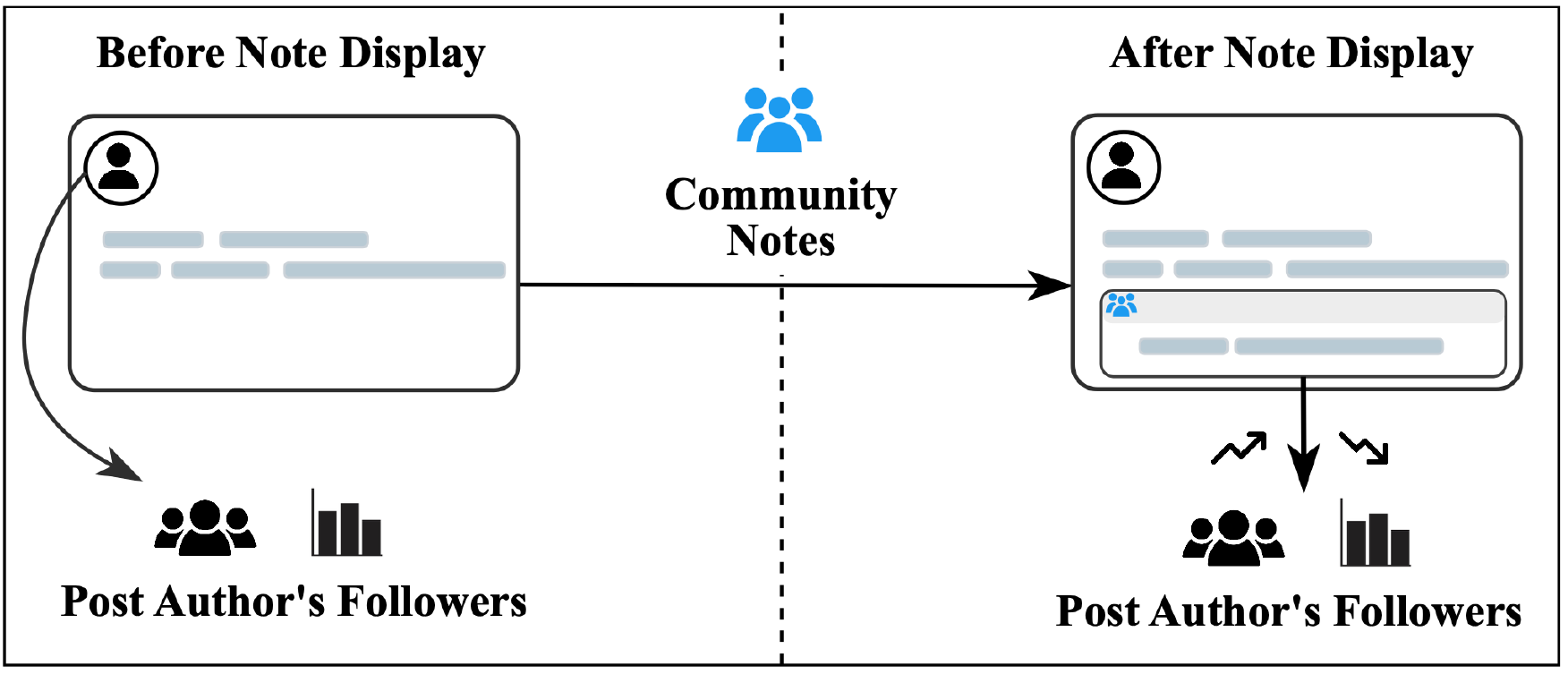}
    \subcaption{} 
    \label{fig:study_sketch}

    \vfill

    \includegraphics[width=\textwidth]{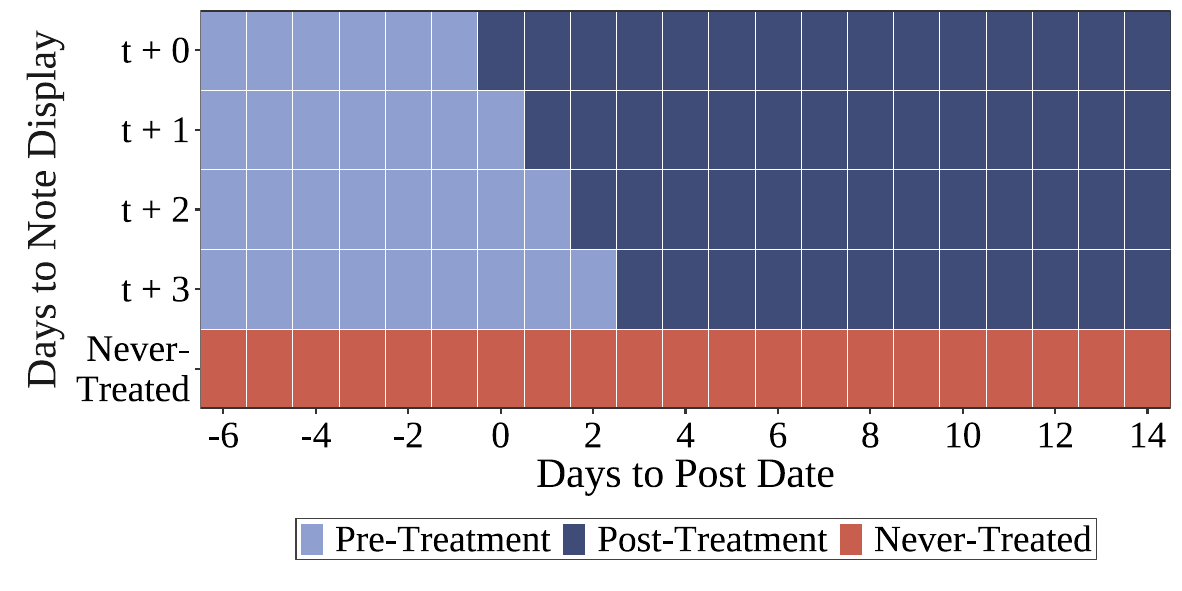}
    \subcaption{} 
    \label{fig:staggered_setup}
\end{minipage}

\caption{\textbf{Study overview.} (a) Example of a misleading post displaying a community note on X. (b) Illustration of the research setup. (c) Visualization of the staggered treatment adoption.}
\label{fig:Test}
\end{figure*}

Effectively countering the spread of misinformation requires fact-checking strategies that are both accurate and scalable. A common approach involves third-party fact-checking organizations, such as \textit{politifact.com} or \textit{snopes.com}, who identify and flag misleading content \cite{Wu.2019,  Chuai.2025b}. While such fact-checks are generally highly accurate, they face significant limitations in speed and scale due to the sheer volume of misinformation produced each day \cite{Micallef.2022, Pennycook.2019}. Previous studies also show that users often distrust third-party fact-checkers, perceiving them as biased \cite{Poynter.2019, Drolsbach.2024}. Another approach that is scalable but suffers from lower accuracy involves machine learning methods \cite{Ma.2016, Wu.2019}.

A growing body of literature explores an alternative approach to combating misinformation: outsourcing fact-checking to platform users themselves \cite{Allen.2021, Bhuiyan.2020, Pennycook.2019, Prollochs.2022a, Drolsbach.2023, Drolsbach.2023b}. This approach builds on the principle of the ``wisdom of crowds'', which posits that the aggregate assessment of a diverse group of non-experts is comparable to that of an expert. Evidence from several studies has shown that this holds even for relatively small groups \cite{Bhuiyan.2020, Epstein.2020, Resnick.2021}. 

A prominent application of community-based fact-checking is the Community Notes \cite{Twitter.2021,Prollochs.2022a} feature on X (initially launched as Birdwatch). Introduced globally in December 2022, this feature allows users to append short contextual information to posts they believe to be misleading or not misleading, using textual annotations of up to \num{280} characters \cite{CNsFAQ.2024}. These annotations, known as community notes, are subsequently rated by other users as helpful or not helpful. A note is only displayed publicly beneath a post if consensus on its helpfulness is reached among a diverse set of contributors (see example in Fig. \ref{fig:note_example}).

Recent research highlights several advantages of community notes. They help scale fact-checking coverage by enabling more posts to be annotated \cite{Pennycook.2019, Chuai.2024}, and users perceive these notes as more trustworthy \cite{Drolsbach.2024}. While concerns have been raised about potential political bias in user-generated notes \cite{Allen.2022, Prollochs.2022a}, there is also evidence that users perceive them as informative, helpful, and tend to agree with their content \cite{Prollochs.2022a, Drolsbach.2023b, Solovev.2025}.

\subsection{Research Gap} 

A growing body of literature has examined the effects of community notes on engagement with misleading content. For instance, \citet{Chuai.2024b} use causal inference techniques to demonstrate that misleading posts with a visible note receive significantly fewer reposts than comparable posts without one. In a similar vein, \citet{Wojcik.2022} show in a study conducted directly on X that the display of a community note can reduce reposts by up to \SI{34}{\percent}. However, these studies focus on content-level outcomes, leaving the consequences for the individuals who authored the fact-checked posts unexplored. Recent evidence by \citet{Kim.2025} reveals that misinformation is often posted when users venture outside their ideological bubbles, and that individual fact-checking can drive users back into echo chambers. In contrast, community notes appear to mitigate these unintended effects. These findings raise critical questions about how community-based fact-checking affects not only what users see but how they are socially perceived and connected. To the best of our knowledge, this is the first study to causally identify the effects of community-based fact-checking on the consequences for users receiving a note.

\section{Data and Methods}

\subsection{Data Sources}

In this study, we causally estimate the effect of displaying community notes on a user's follower base. To this end, we collect data from three sources: (i) community notes from X, (ii) the underlying fact-checked posts, and (iii) daily follower data from Social Blade. 

\textbf{Community Notes dataset:} 
X provides daily updates on all community notes and their status histories on a dedicated website\footnote{https://x.com/i/communitynotes/download-data}. From this dataset, we select all English-language community notes for posts flagged as misleading between March 1, 2024, and August 29, 2024, \ie, for a period of six months. This includes both notes rated as helpful (\ie, those that have been publicly displayed), and those that remained invisible (\ie, never received the {helpful} status), which serve as controls in our empirical design. Given that posts receive on average \num{1.26} notes, we retain the first note rated as helpful, or the earliest authored note for posts that never displayed a note \cite{Drolsbach.2023b}. This yields a dataset of \num{190873} individual notes, each corresponding to a distinct post. 

\begin{figure}[t]
  \begin{subfigure}{0.49\columnwidth}
  \includegraphics[width=\textwidth]{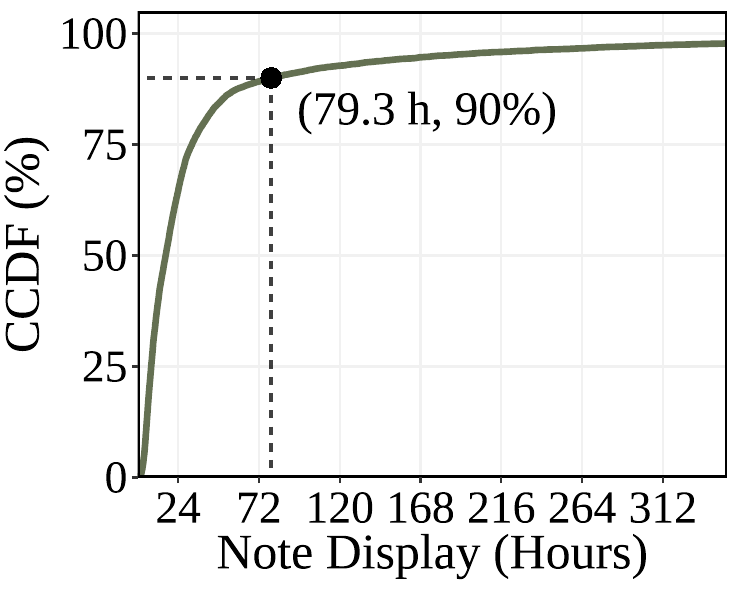}
  \caption{}
    \label{fig:notes_display}
  \end{subfigure}
  \hfill
  \begin{subfigure}{0.49\columnwidth}
  \includegraphics[width=\textwidth]{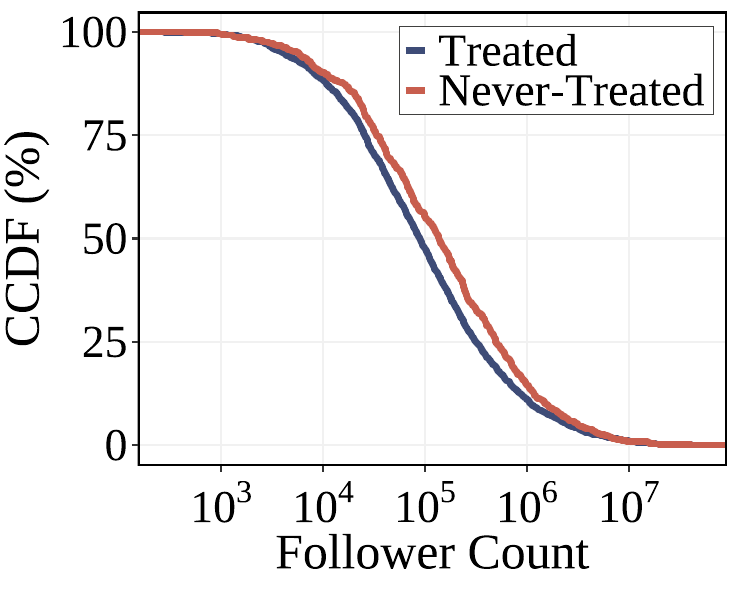}
  \caption{} 
     \label{fig:abs_follower_count}
  \end{subfigure} 
  \begin{subfigure}{0.49\columnwidth} 
  \includegraphics[width=\textwidth]{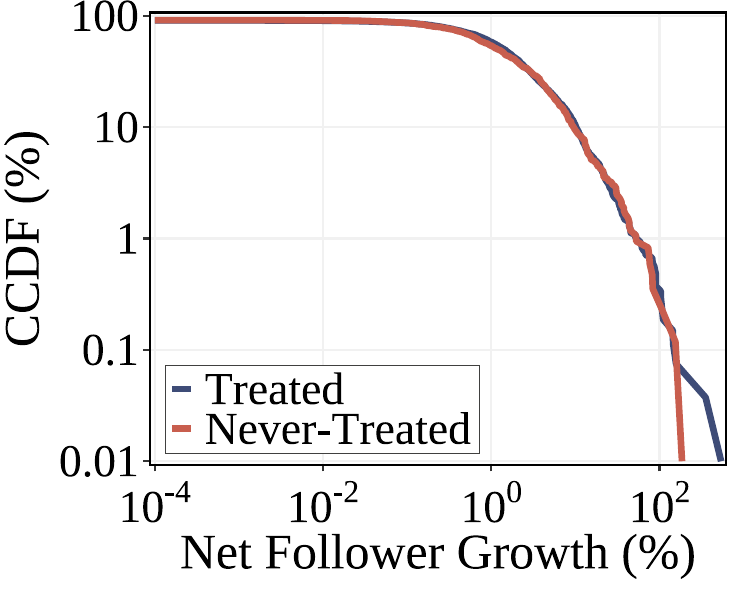} 
  \caption{} 
     \label{fig:net_growth}
  \end{subfigure}  
  \hfill 
  \begin{subfigure}{0.49\columnwidth} 
  \includegraphics[width=\textwidth]{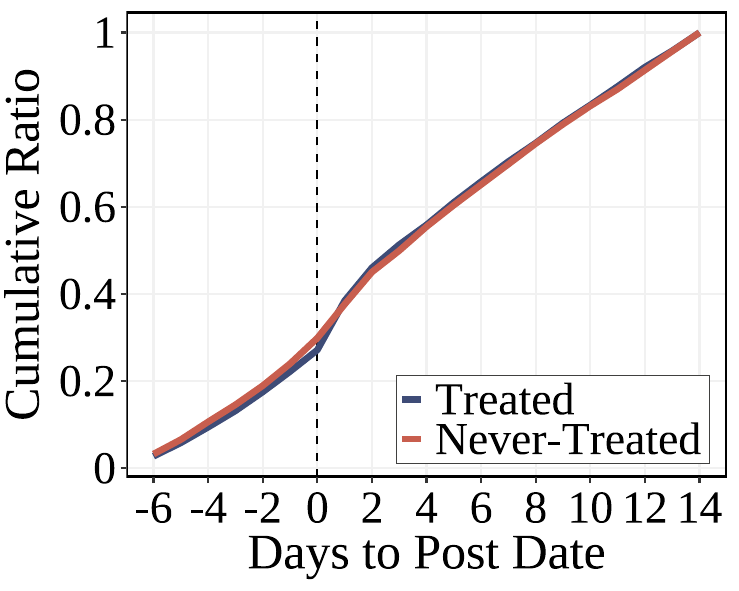} 
  \caption{} 
     \label{fig:follower_growth}
  \end{subfigure}
  \caption{\textbf{Descriptive Statistics of Note Display and Followers.} (a) Complementary cumulative distribution function (CCDF) of note display timing, showing that most notes become visible within three days of post publication. (b) CCDF of absolute follower count for treated and never-treated users. (c) CCDF of net follower growth during the 21-day observation window, showing that most users experience minimal change in follower counts. (d) Normalized cumulative follower growth, comparing the growth trajectories over the observation period for treated and never-treated units.}
  \end{figure}

\textbf{Post dataset:} 
Using the post IDs from the community notes dataset, we retrieve metadata on both the post and its author via the X API v2 lookup endpoint. To manage API costs, we focus on a random sample of \num{24000} posts from the notes dataset. To improve covariate balance and reduce the influence of extreme observations, we apply propensity score weights based on user and post characteristics and trim users in the tails of the distribution. We restrict the sample to users with more than \num{500} followers to increase the probability of users being tracked on Social Blade, which primarily monitors larger content creators. While some users are noted more frequently than others, only about \SI{1.8}{\percent} in our sample receive more than ten notes. To mitigate potential skew from these users, we also filter them out. Finally, since approximately \SI{90}{\percent} of \emph{helpful} notes are displayed within three days (see Fig.~\ref{fig:notes_display}), we focus on notes shown during this period. This restriction also ensures that we capture follower behavior during the critical window of peak engagement. After applying these steps, our dataset comprises \num{13083} posts from \num{6508} different users.

\textbf{Follower data from Social Blade:}
Since the X API does not provide access to historical follower data, we collect historical daily follower counts from Social Blade (\url{https://socialblade.com/}), which tracks public metrics for a wide range of social media accounts. To balance API costs, we randomly sampled \num{4250} user accounts from our filtered post dataset. Given API limitations and some users not being tracked on the platform, we were able to retrieve follower data for \num{2142} users. 
Subsequently, we merge the different data sources to construct a longitudinal dataset at the post-day level. To ensure comparability across observations, we center all posts around their publication date and restrict the observation window to seven days before and 14 days after the post goes online, an event common to both treated and never-treated posts. Although the sample includes posts authored until the end of August, the observation period extends into September to accommodate the full 21-day tracking period. The final dataset comprises \num{73836} observations, covering \num{3516} unique posts authored by \num{2142} accounts.

\subsection{Key Variables}

\textbf{Dependent variable:}
Our main outcome variable is the daily percentage change in follower count, calculated as the log difference between consecutive days. This specification captures immediate shifts in follower behavior in response to the display of a note. Given that follower counts are highly right-skewed \cite{Kivran-Swaine.2011, Kwak.2011}, this transformation allows for better comparability across accounts of vastly different sizes and stabilizes variance. 

\textbf{Explanatory variables:}
We collected the following user and post characteristics. These serve as explanatory variables in our later empirical analysis and allow us to match similar treated and not-yet-treated units.

\smallskip
\underline{User-level variables}:
\begin{itemize}
    \item \emph{Account Age}: The number of years since a user created his/her account on X. 
    \item \emph{\#Posts}: The number of posts a user has posted since account creation.
    \item \emph{\#Followers}: The number of accounts that follow a user. 
    \item \emph{\#Followees}: The number of accounts that a user follows.
    \item \emph{Verified}: A dummy indicating whether X has officially verified a user ($= 1$; 0 otherwise). 
\end{itemize}

\smallskip
\underline{Post-level variables:}
\begin{itemize}
    \item \emph{\#Reposts}: The number of times the post was reposted by other users. 
    \item \emph{\#Replies}: The number of replies made to the source post.
    \item \emph{Media}: A binary indicator of whether a source post includes media, such as an image or video ($= 1$; 0 otherwise). 
   \item \emph{\#Words}: We remove user mentions, URLs, convert HTML to Unicode to then apply ICU breakiterators to count the number of words per post. 
   \item \emph{Sentiment}: We calculate sentiment scores \cite{Feuerriegel.2025} using the Twitter-roBERTa-base model \cite{Loureiro.2022}, and classify posts as \emph{positive} or \emph{negative} given the highest predicted probability. 
   \item \emph{Political}: Using X's post annotations \footnote{https://docs.x.com/x-api/fundamentals/post-annotations}, we classify posts as \emph{political} based on the keywords ``Politician'', ``Political Race'', and ``Political Body''.
\end{itemize}

Although the X API provides various engagement metrics (\eg, likes, reposts, replies, and quotes), we focus on reposts and replies due to their high correlation \cite{Pilarski.2024}, which helps mitigate multicollinearity affecting the estimation.

\subsection{Model Specification}

In recent years, a substantial body of literature has highlighted limitations of the commonly used two-way fixed effects (TWFE) estimator in settings with staggered treatment adoption. Specifically, TWFE can yield biased estimates due to inappropriate comparisons between already-treated and newly-treated units \cite{deChaisemartin.2020, GoodmanBacon.2021, Sun.2021}. To address these concerns, we employ the estimator proposed by \citet{Callaway.2021}, which is tailored to staggered difference-in-differences (DiD) designs by explicitly accounting for treatment effect heterogeneity across groups treated at different times.

The core idea behind this approach is to group treated units based on when they first receive treatment and to estimate average treatment effects by comparing their outcomes to units that are never treated or treated at a later point in time (not-yet-treated). In our case, we use the not-yet-treated group as comparison group, which includes both posts that are never treated and those who have not yet received a note at time $t$.

To provide intuition for the empirical setup, we consider the following stylized model: 
\begin{align}
Y_{ijt} = X'\beta + \tau_{it} \cdot Display_{it} + \lambda_{ij} + \mu_t + \varepsilon_{ijt},
\label{eq:stylized_reg}
\end{align}

where $Y_{ijt}$ is the daily percentage change in followers for user $j$ relative to post $i$, $X_{ijt}$ represents a vector of our time-invariant post-level and user-level covariates, $Display_{it}$ is a binary indicator if a post is treated (\ie, displayed) at time $t$, $\tau_{it}$ captures the corresponding treatment effect, $\lambda_{ij}$ represents post-user fixed effects, and $\mu_t$ captures day fixed effects, which absorb systematic trends across posts, allowing us to isolate the treatment effect from general post-life-cycle trends. While the actual estimation procedure follows a two-step approach to obtain group-time average treatment effects \cite{Callaway.2021}, the model provides intuition for the core identification logic. 

Formally, the estimator constructs group-time average treatment effects using the following expression: 
\begin{align}
ATT(g,t) &= \mathbb{E}[Y_t - Y_{g-1} \mid G_g = 1] \nonumber \\
         &\quad - \mathbb{E}[Y_t - Y_{g-1} \mid D_t = 0, G \neq 1],
\label{eq:group_time_att}
\end{align}

where \emph{g} denotes the day relative to post publication on which a post first receives treatment, and $D_t=0$ identifies units untreated at time \emph{t}. In our setting, posts are first treated on days $+0$ to $+3$ following publication. As illustrated in Fig.~\ref{fig:staggered_setup}, some notes become visible immediately ($t + 0$), while others are treated one ($t + 1$), two ($t + 2$), or three days ($t + 3$) later. Posts labeled ``never-treated'' received a note, but they never passed the \textit{helpful}-threshold and thus remained invisible to the public. All posts are observed over a 21-day window. This staggered adoption of treatment motivates the use of a difference-in-differences framework accounting for variation in treatment timing, comparing changes in outcomes for treated and note-yet-treated posts. 

We implement the doubly robust version of the estimator, which combines outcome regression with inverse probability weighting based on our set of post- and user-level covariates. Under the assumptions that (i) treated and untreated units would have followed similar trends in the absence of treatment (conditional parallel trends), and (ii) treatment effects do not occur before a note is displayed (no anticipation), the group-time ATTs can be interpreted causally \cite{Callaway.2021}.

The group-time ATTs derived from Equation ~\ref{eq:group_time_att} are further aggregated across treatment days and event times to summarize treatment effects by treatment timing and over time. This facilitates an intuitive interpretation of both heterogeneity in treatment exposure and temporal dynamics. To address correlation in the outcome across time and users, standard errors are clustered at the user-post level. Furthermore, to enable valid inference, we report simultaneous confidence bands based on \num{5000} bootstrap replications, which account for multiple testing concerns \cite{Callaway.2021}.

\section{Empirical Analysis}

In this section, we empirically analyze the causal effect of community notes on follower counts (see study overview Fig.~\ref{fig:study_sketch}). We begin by providing an overview of our dataset and key descriptive patterns (full descriptive statistics are in the SI, Table \ref{tab:data_summary}). Subsequently, we report our main causal estimates, followed by a wide variety of checks and sensitivity analyses to validate the robustness of our findings. Unless otherwise stated, all analyses report 95\% confidence intervals (abbreviated as ‘CI’).

\subsection{Data Overview}

Our longitudinal dataset contains \num{3516} posts from \num{2142} unique users, observed over a \num{21}-day window spanning from one week prior to post publication to two weeks after. Treated posts account for approximately \SI{76}{\percent} of all observations, with the remaining \SI{24}{\percent} never displayed a helpful note. Among the treated posts, \SI{40.1}{\percent} receive a helpful note one day after post publication. The rest of the treated post receive their helpful note on the day of post publication (\SI{22.5}{\percent}), two days (\SI{9.9}{\percent}), or three days (\SI{3.5}{\percent}) later. 

\begin{figure*}
  \begin{subfigure}{0.67\textwidth}
  \includegraphics[width=\linewidth]{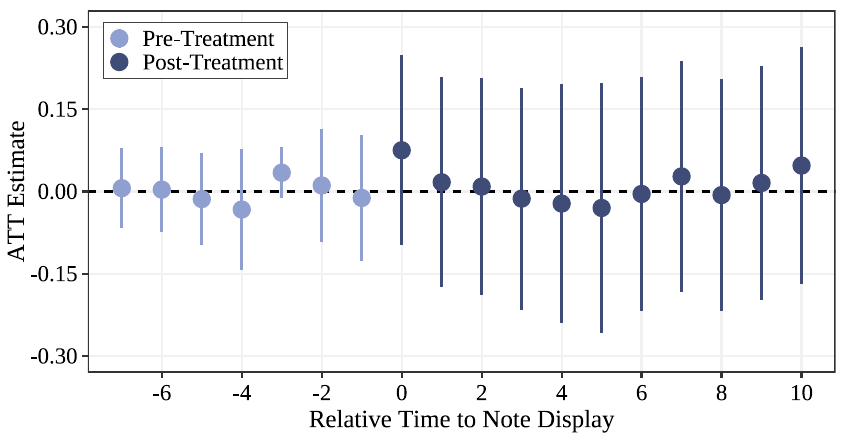}
  \caption{}
  \label{subfig:event_atts_main_spec}
  \end{subfigure}
  \hfill
  \begin{subfigure}{0.32\textwidth}
  \includegraphics[width=\linewidth]{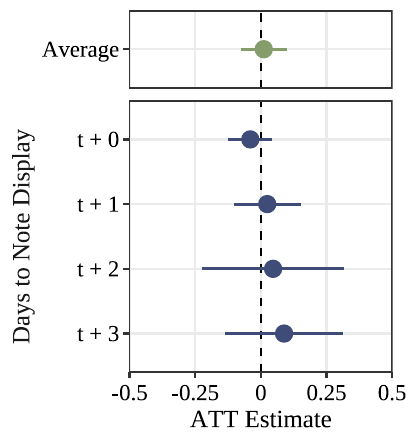}
  \caption{} 
  \label{subfig:group_atts_main_spec}
  \end{subfigure} 
  \caption{\textbf{Estimation results.} Shown are coefficient estimates (circles) and their \SI{95}{\percent} simultaneous confidence intervals (bars) for (a) event-study estimates spanning seven days before and \num{14} after note display, and (b) group-level aggregated estimates, including the simple weighted average across groups.}
  \label{fig:atts_main_spec}
\end{figure*}

\textbf{Baseline differences:} Fig.~\ref{fig:abs_follower_count} compares the absolute follower counts of treated vs. never-treated users. Never-treated users generally exhibit higher absolute follower counts ($\mu_{treated} =$ \num{799273.34}, $\mu_{never} =$ \num{981490.53}, [KS-test: $D=0.098$, $p<0.001$]). However, when comparing relative follower growth, the distributions are more aligned. Fig.~\ref{fig:net_growth} displays the complementary cumulative distribution function of total follower gain from day 1 to day 21, normalized by each user’s baseline count.  The distributions of treated and never-treated users follow a similar trajectory, with some small divergence in the upper tails. Although the difference in distributions is marginally statistically significant, the practical relevance is likely limited $\mu_{treated} =$ \SI{4.71}{\percent}, $\mu_{never} =$ \SI{4.51}{\percent}, [KS-test: $D=0.054$, $p=0.045$]). Notably, \SI{75}{\percent} of users gain fewer than \SI{4}{\percent} new followers over the entire observation window, indicating that follower counts tend to remain relatively stable overall. 

\textbf{Temporal follower dynamics:} While baseline differences offer important context, they do not reveal the fluctuations in follower growth over time. Fig.~\ref{fig:follower_growth} shows the normalized follower growth, where each user's total growth over the 21-day window is scaled from $0$ to $1$. This makes it possible to directly compare how quickly treated and never-treated users accumulate followers relative to the post date. Both groups follow broadly similar growth trajectories, though treated users exhibit a slightly sharper increase immediately after publication. Still, follower counts in levels appear inert to treatment: the average log follower count on the day immediately before and after note display is nearly identical ($\mu_{before} = 11.702$, $\mu_{after}=11.709$, [t-test: $t = 0.145$, $p=0.885$]). Similarly, there is no statistically significant difference when considering a week before and after note display ($\mu_{before} = 11.701$, $\mu_{after}=11.703$, [t-test: $t = 1.480$, $p=0.911$]). This suggests that follower counts in levels may be too stable to detect short-run effects. 

 In contrast, daily growth rates provides a more sensitive measure. Treated users grow significantly faster following note display ($\mu_{before} = 0.269\%$, $\mu_{after}=0.423\%$, [t-test: $t = 4.297$,  $p<0.001$]). However, this spike coincides with a broader pattern seen across all users: both treated and never-treated users experience increased follower growth immediately after post publication. For treated users, growth increases from \SI{0.24}{\percent} to \SI{0.37}{\percent} (t-test: $t = 3.687$, $p<0.001$), whereas never-treated users experience an increase from \SI{0.19}{\percent} to \SI{0.30}{\percent} (t-test: $t = 2.824$, $p<0.001$). This suggest that post-driven exposure, rather than fact-checking itself, may drive much of the follower dynamics -- and highlights the need for causal strategies to disentangle the effect of a community note from organic follower growth.

\subsection{Causal Effect of Community Notes on Followers}

We now report the estimates of our DiD model estimating the causal effect of community notes on follower growth.
Fig.~\ref{subfig:event_atts_main_spec} shows the estimated ATTs aggregated by event-time, that is, for each day relative the display of a community note. The analysis covers a 21-day window, spanning from seven days before to \num{14} days after treatment. 

\textbf{Pre-treatment effects: }We observe no statistically significant anticipatory effects: all pre-treatment estimates are centered around zero, supporting the parallel trends assumption. This is further supported by a pre-trends test based on conditional moments \cite{Callaway.2021}, which yields a $p$-value of \num{0.5591}, suggesting no significant differences in trends between treated and not-yet-treated units prior treatment. 

\textbf{Post-treatment effect: }
On the day of treatment, we observe a small increase in follower growth of 0.07 percentage points that is, however, not statistically significant at the 95\% statistical significance level (\SI{95}{\percent} CI: [\num{-0.10}, \num{0.25}]). In the days following treatment, the ATT estimates remain similarly small in magnitude and are also not statistically significant at conventional significance levels. Overall, this implies that the display of community notes does not meaningfully affect follower growth. 

\textbf{Group-level effects: }
To examine heterogeneity by treatment timing, we aggregate our estimates for each treatment group (see Fig. \ref{subfig:group_atts_main_spec}). Across all treatment days, we find estimates close to zero that are not significant at common statistical significance levels. The most pronounced (but still small) effect is observed for posts treated three days after post publication with a magnitude of \num{0.09} percentage points, though the effect is not statistically significant (\SI{95}{\percent} CI: [\num{-0.29}, \num{0.46}]). The average ATT across all treatment groups, likewise not statistically significant, is \num{0.01} percentage points (\SI{95}{\percent} CI: [\num{-0.14}, \num{0.16}]). Overall, these results reinforce our main finding that community notes do not lead to meaningful declines in followers.

\subsection{Robustness checks}

To validate the robustness of our findings, we conduct a series of supplementary analyses using alternative model specifications and sample restrictions. Specifically, we (i) estimate the model without covariates, (ii) exclude the top \SI{10}{\percent} most volatile users, \ie, those experiencing the most pronounced changes in average daily follower growth, (iii) restrict the sample to users who are treated exactly once in our observation period to address concerns about repeated treatments, and (iv) use an alternative control group consisting of only never-treated posts (see SI, Table \ref{tab:group_att_growth_rate} and SI, Table \ref{tab:event_time_att_growth_rate}). Across all checks, the ATTs remain close to zero and statistically not significant, reinforcing the conclusion that community notes have no meaningful effect on follower growth. 

\textbf{Analysis of cumulative changes:} To complement our main outcome measure based on daily growth rates, we replicate the analyses using follower counts expressed in logs, capturing cumulative changes rather than day-to-day variation. The results are consistent with those from the models utilizing the daily follower growth rate as outcome: all event-time ATTs remain small in magnitude and statistically not significant (see SI, Table \ref{tab:group_att_follower_count} and SI, Table \ref{tab:event_time_att_follower_count}). One exception emerges when we restrict the sample to users treated exactly once, yielding a marginally significant reduction in followers of \SI{0.66}{\percent} for the latest treated group in our setup (ATT = \num{-0.0066}, \SI{95}{\percent} CI: [\num{-0.0121}, \num{-0.0012}]). However, this pattern does not replicate across the other model specifications or sub-samples. 

\subsection{Sensitivity Analysis}

\begin{figure}[t]
  \begin{subfigure}{0.58\columnwidth} 
  \includegraphics[width=\linewidth]{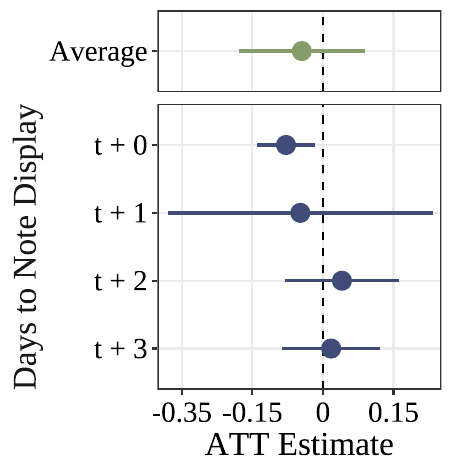} 
  \caption{} 
     \label{fig:att_large_accounts_growth}
  \end{subfigure}  
  \hfill 
  \begin{subfigure}{0.4\columnwidth} 
  \includegraphics[width=\linewidth]{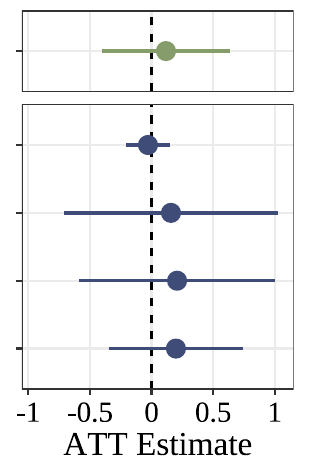} 
  \caption{} 
     \label{fig:att_small_accounts_growth}
  \end{subfigure}
  
  \begin{subfigure}{0.58\columnwidth} 
  \includegraphics[width=\linewidth]{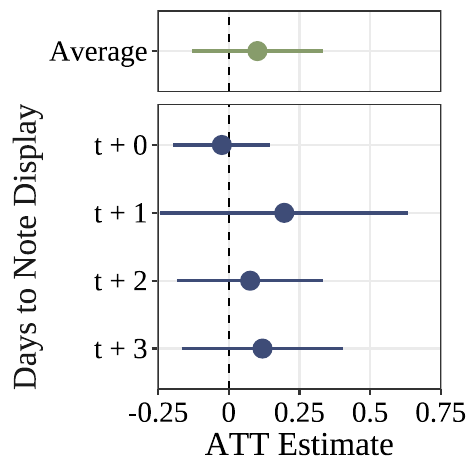} 
  \caption{} 
     \label{fig:att_political_growth}
  \end{subfigure}  
  \hfill 
  \begin{subfigure}{0.4\columnwidth} 
  \includegraphics[width=\linewidth]{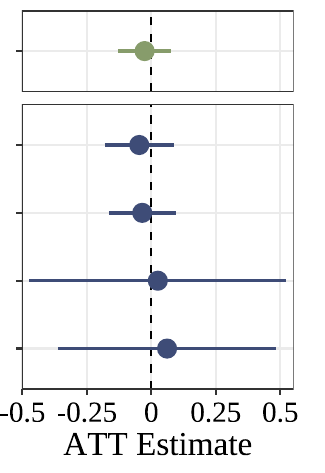} 
  \caption{} 
     \label{fig:att_non_political_growth}
  \end{subfigure}
  \caption{\textbf{Estimation Results of Sensitivity Analyses.} Shown are coefficients (circles) and their \SI{95}{\percent} simultaneous confidence intervals (bars) for distinct sub-samples, namely (a) large accounts, i.e., users who have a more followers than the sample median, (b) small accounts, i.e., accounts with follower counts below the sample median, (c) posts containing political content, and (d) post containing non-political content.}
\end{figure}

To rule out the possibility that opposing treatment effects for different posts may cancel each other out when aggregated, we split our sample into subgroups and re-estimate our models. We evaluate both daily follower growth and log follower counts as outcome variables, on sub-samples split by account size (small vs. large accounts) and topic (political vs. non-political posts). 

\textbf{Analysis across small vs. large accounts:} We divide the sample at the median follower count to distinguish between small and large accounts. For large accounts, \ie, for users whose follower count exceeds the sample median, we find a statistically significant negative effect of \num{0.08} percentage points (\SI{95}{\percent} CI: [\num{-0.14}, \num{-0.02}]) in daily follower growth when the community note appears on the same day as the post is published (see Fig. ~\ref{fig:att_large_accounts_growth}). However, this effect is not robust when we use the number of followers as outcome variable. Nonetheless, when using the alternative outcome of the absolute follower count we find a drop of about \num{0.47} percentage points for posts treated on day 3 following note display (ATT = \num{-0.0047}, \SI{95}{\percent} CI: [\num{-0.0093}, \num{-0.0002}]).  Fig.~\ref{fig:att_small_accounts_growth} reveals no significant effects for small accounts, although this sub-sample yields the largest aggregated ATT in terms of effect size (ATT = \num{0.12}, \SI{95}{\percent} CI: [\num{-0.13}, \num{0.33}]).

\textbf{Analysis across political vs. non-political posts:} We further split our sample based on whether a post is labeled as political, using X’s content annotations. We do not find statistically significant effects for either sub-sample using both outcome specifications, as can be taken from Fig.~\ref{fig:att_political_growth} and Fig.~\ref{fig:att_non_political_growth}. Political posts react positively (ATT = \num{0.10}, \SI{95}{\percent} CI: [\num{-0.13}, \num{0.33}]), whereas non-political post exhibit a small but negative aggregated ATT (ATT = \num{-0.03}, \SI{95}{\percent} CI: [\num{-0.13}, \num{0.07}]), when using the daily follower growth as dependent variable. 

\section{Discussion}


Community-based fact-checking is increasingly adopted by social media platforms \cite{Youtube.2024, Meta.2025, TikTok.2025} as a faster, more scalable, and more trusted approach to addressing misinformation \cite{Pennycook.2019, Drolsbach.2024}. While prior work demonstrates that community-based fact-checking reduces engagement with misleading content once it is flagged \cite{Chuai.2024b}, evidence on how they affect the followers of users who are fact-checked is missing. This study addresses this gap by empirically analyzing how community notes on X influence follower dynamics. Using quasi-experimental methods, we provide \emph{causal} evidence that being community-noted does not lead to a statistically significant decline in follower counts. 

\subsection{Research Implications} 

Our study identifies a key limitation of community-based fact-checking: its inability to disrupt the social followership that sustains misinformation. Across a wide range of model specifications, outcome measures, and sub-samples, we find no consistent or measurable effect of community notes on follower dynamics. This suggests that even when misinformation is publicly corrected, the reputational consequences for the author may be minimal.

One possible explanation for our findings is that community notes might primarily influence \emph{non-followers}, \ie, users who encounter the post via indirect exposure. For these users, the presence of a community note note may discourage engagement with the content \cite{Chuai.2024b}. In contrast, a fact-checked user's \emph{followers} may remain loyal because their relationship with the user is driven more by social or ideological factors \cite{Aiello.2012, Barbera.2015a} rather than content credibility \cite{Ashkinaze.2024}. This interpretation aligns with evidence from prior research showing that while fact-checks can reduce beliefs in false claims, they often fail to change attitudes towards the person who authored it, particularly in political contexts \cite{Swire-Thompson.2020, Nyhan.2020}. 

A second potential factor is the timing of community notes' visibility. Prior research shows that half of a post's impressions occur within \num{80} minutes after publication \cite{Pfeffer.2023}. In contrast, the median time for a note to become visible is \num{16} hours in our dataset. This suggests that notes often appear after a post experienced peak engagement, implying that many users might never encounter the corrective note, thereby limiting it's potential to meaningfully alter follower behavior. 

Our findings align well with observational evidence from prior work on conversational fact-checking, which found that misinformation spreaders are rarely unfollowed when corrected via replies linking to professional fact-checkers \cite{Ashkinaze.2024}. However, they contrast with survey-based evidence suggesting that users \emph{intend} to unfollow or block those who share misinformation \cite{Kaiser.2022}. By providing real-world causal evidence in the context of community-based fact-checking, our study offers new insights into the limits of corrective interventions -- and highlights a gap between what users say they would do when confronted with spreaders of misinformation and their actual behavior on social media.

While the overall effect of community notes on follower dynamics is not statistically significant, we do observe small but significant effects within specific subgroups. Specifically, large accounts receiving a note on the same day as the post being authored, experience a short-term decline in follower growth, though this effect does not translate into a longer-term follower loss. In contrast, posts flagged three days after post publication exhibit cumulative negative effects on follower counts in two sub-samples: (i) large accounts, and (ii) users treated exactly once. One possible reason for this pattern could be that later notes reintroduce the post to public attention, \eg via reposts or algorithmic resurfacing. These patterns may also reflect the social nature of follower dynamics. Prior research shows that users are less likely to break reciprocal ties or unfollow similar users \cite{Xu.2013, Hutto.2013, Kwak.2011, Kivran-Swaine.2011}. For large accounts, where social ties are more likely to be parasocial, the perceived social cost of unfollowing could be lower. Consequently, users may be more willing to disengage from such accounts. Still, even within these subgroups, the effect sizes are small, reinforcing the conclusion that the influence of community notes on followers is minimal.

\subsection{Practical Implications}

From a practical standpoint, our findings suggests community-based fact-checking is unlikely to address the deeper social networks and reputational dynamics that sustain misinformation. In particular, we find that it is insufficient to disrupt follower relationships with spreaders of misinformation, which are often rooted in ideological alignment or social affinity \cite{Aiello.2012, Barbera.2015a} rather than content credibility \cite{Ashkinaze.2024}. This highlights the need for complementary strategies that target not only misleading content, but also the social connections through which it spreads. This challenge is especially relevant for platforms like X, which rely predominantly on community-based moderation \cite{Drolsbach.2024b, Trujillo.2025, Kaushal.2024}. 

To strengthen the impact of community-based fact-checking, platforms could implement additional strategies such as down-ranking accounts that are repeatedly fact-checked, issuing prompts that encourage users to reconsider following misinformation sources, or introducing transparency features that disclose a user's history of being fact-checked. Beyond platform-level interventions, public education efforts aimed at improving media literacy and critical consumption of social media content may help foster greater follower-level accountability.

\subsection{Limitations \& Future Research}

As with any research, our study has limitations that suggest promising directions for future work. First, our analysis is limited to a single platform (X) and content in English. Future research could explore whether similar patterns hold across other platforms or cultural settings. Second, our 21-day observation window captures short-term effects but may miss longer-term user responses. Extending the timeframe could help evaluate whether repeated fact-checks gradually erode audience loyalty. Third, although our sample is well-suited for causal identification, its moderate size limits our ability to detect very small effects. However, the fact that estimated effects are consistently near zero across models suggests that any true effects are likely minimal in practical terms.  Fourth, we rely on aggregate follower counts and cannot track which specific users choose to follow or unfollow. Future studies with access to network-level data could offer a richer view of how fact-checking influences specific user segments. Finally, while we focus on the effects for users who are fact-checked, future work could examine behavioral responses on the part of those users themselves -- such as changes in the volume, misleadingnesss, or tone of their subsequent content. 

\section{Conclusion}

This study provides causal evidence that community-based fact-checking on social media does not lead to a statistically significant decline in the follower counts of users who are fact-checked. While prior work has demonstrated that community notes can reduce engagement (\eg, likes, reposts) with misleading content, we find that the authors of such content generally retain their followers. Even subgroup analyses reveal only minimal, short-lived effects, primarily among large accounts. Taken together, our findings suggest that while community-based fact-checking can curb content-level engagement, it may be insufficient to disrupt the social ties that help misinformation persist. As platforms increasingly rely on community-based fact-checking systems, our results highlight the need for complementary strategies to more effectively counteract misinformation on social media.

\section{Ethics Statement}
All analyses are based on publicly available data. The data collection and the analysis follow common standards for ethical research \cite{Rivers.2014}. We declare no competing interests.

\bibliography{references}

\begin{thebibliography}{81}
\providecommand{\natexlab}[1]{#1}

\bibitem[{Aiello et~al.(2012)Aiello, Barrat, Schifanella, Cattuto, Markines,
  and Menczer}]{Aiello.2012}
Aiello, L.~M.; Barrat, A.; Schifanella, R.; Cattuto, C.; Markines, B.; and
  Menczer, F. 2012.
\newblock Friendship prediction and homophily in social media.
\newblock \emph{ACM Transactions on the Web}, 6(2): 1--33.

\bibitem[{Allcott and Gentzkow(2017)}]{Allcott.2017}
Allcott, H.; and Gentzkow, M. 2017.
\newblock Social media and fake news in the 2016 election.
\newblock \emph{Journal of Economic Perspectives}, 31(2): 211--236.

\bibitem[{Allen et~al.(2021)Allen, Arechar, Pennycook, and Rand}]{Allen.2021}
Allen, J.; Arechar, A.~A.; Pennycook, G.; and Rand, D.~G. 2021.
\newblock Scaling up fact-checking using the wisdom of crowds.
\newblock \emph{Science Advances}, 7(36): eabf4393.

\bibitem[{Allen, Martel, and Rand(2022)}]{Allen.2022}
Allen, J.; Martel, C.; and Rand, D.~G. 2022.
\newblock Birds of a feather don’t fact-check each other: Partisanship and
  the evaluation of news in Twitter’s Birdwatch crowdsourced fact-checking
  program.
\newblock In \emph{CHI}.

\bibitem[{Altay, Hacquin, and Mercier(2020)}]{Altay.2020b}
Altay, S.; Hacquin, A.-S.; and Mercier, H. 2020.
\newblock {Why do so few people share fake news? It hurts their reputation}.
\newblock \emph{New Media {\&} Society}, 24(6): 1303--1324.

\bibitem[{Ashkinaze, Gilbert, and Budak(2024)}]{Ashkinaze.2024}
Ashkinaze, J.; Gilbert, E.; and Budak, C. 2024.
\newblock The Dynamics of (Not) Unfollowing Misinformation Spreaders.
\newblock In \emph{WWW}.

\bibitem[{Bakshy et~al.(2011)Bakshy, Hofman, Mason, and Watts}]{Bakshy.2011}
Bakshy, E.; Hofman, J.~M.; Mason, W.~A.; and Watts, D.~J. 2011.
\newblock Everyone's an influencer.
\newblock In \emph{WSDM}.

\bibitem[{Bakshy, Messing, and Adamic(2015)}]{Bakshy.2015}
Bakshy, E.; Messing, S.; and Adamic, L.~A. 2015.
\newblock Exposure to ideologically diverse news and opinion on {F}acebook.
\newblock \emph{Science}, 348(6239): 1130--1132.

\bibitem[{B{\"a}r, Pr{\"o}llochs, and Feuerriegel(2023)}]{Bar.2023}
B{\"a}r, D.; Pr{\"o}llochs, N.; and Feuerriegel, S. 2023.
\newblock New threats to society from free-speech social media platforms.
\newblock \emph{Communications of the ACM}, 66(10): 37--40.

\bibitem[{B{\"a}r, Pröllochs, and Feuerriegel(2023)}]{Baer.2022}
B{\"a}r, D.; Pröllochs, N.; and Feuerriegel, S. 2023.
\newblock Finding Qs: Profiling QAnon supporters on Parler.
\newblock In \emph{ICWSM}.

\bibitem[{Barber{\'a}(2015)}]{Barbera.2015a}
Barber{\'a}, P. 2015.
\newblock {Birds of the Same Feather Tweet Together: Bayesian Ideal Point
  Estimation Using Twitter Data}.
\newblock \emph{Political Analysis}, 23(1): 76--91.

\bibitem[{Barber{\'a} et~al.(2015)Barber{\'a}, Jost, Nagler, Tucker, and
  Bonneau}]{Barbera.2015}
Barber{\'a}, P.; Jost, J.~T.; Nagler, J.; Tucker, J.~A.; and Bonneau, R. 2015.
\newblock Tweeting from left to right: Is online political communication more
  than an echo chamber?
\newblock \emph{Psychological Science}, 26(10): 1531--1542.

\bibitem[{Bhuiyan et~al.(2020)Bhuiyan, Zhang, Sehat, and Mitra}]{Bhuiyan.2020}
Bhuiyan, M.~M.; Zhang, A.~X.; Sehat, C.~M.; and Mitra, T. 2020.
\newblock Investigating differences in crowdsourced news credibility
  assessment: Raters, tasks, and expert criteria.
\newblock In \emph{CSCW}.

\bibitem[{Callaway and Sant{'}Anna(2021)}]{Callaway.2021}
Callaway, B.; and Sant{'}Anna, P. H.~C. 2021.
\newblock Difference-in-Differences with multiple time periods.
\newblock \emph{Journal of Econometrics}, 225(2): 200--230.

\bibitem[{Chuai et~al.(2024{\natexlab{a}})Chuai, Pilarski, Renault,
  Restrepo-Amariles, Troussel-Cl{\'e}ment, Lenzini, and
  Pr{\"o}llochs}]{Chuai.2024b}
Chuai, Y.; Pilarski, M.; Renault, T.; Restrepo-Amariles, D.;
  Troussel-Cl{\'e}ment, A.; Lenzini, G.; and Pr{\"o}llochs, N.
  2024{\natexlab{a}}.
\newblock Community-based fact-checking reduces the spread of misleading posts
  on social media.
\newblock \emph{arXiv}.

\bibitem[{Chuai et~al.(2024{\natexlab{b}})Chuai, Tian, Pr{\"o}llochs, and
  Lenzini}]{Chuai.2024}
Chuai, Y.; Tian, H.; Pr{\"o}llochs, N.; and Lenzini, G. 2024{\natexlab{b}}.
\newblock Did the Roll-Out of {Community Notes} Reduce Engagement With
  Misinformation on {X/T}witter?
\newblock In \emph{CSCW}.

\bibitem[{Chuai et~al.(2025)Chuai, Zhao, Pr{\"o}llochs, and
  Lenzini}]{Chuai.2025b}
Chuai, Y.; Zhao, J.; Pr{\"o}llochs, N.; and Lenzini, G. 2025.
\newblock Is Fact-Checking Politically Neutral? Asymmetries in How US
  Fact-Checking Organizations Pick Up False Statements Mentioning Political
  Elites.
\newblock In \emph{ICWSM}.

\bibitem[{de~Chaisemartin and D'Haultf{\oe}uille(2020)}]{deChaisemartin.2020}
de~Chaisemartin, C.; and D'Haultf{\oe}uille, X. 2020.
\newblock {Two-Way Fixed Effects Estimators with Heterogeneous Treatment
  Effects}.
\newblock \emph{American Economic Review}, 110(9): 2964--96.

\bibitem[{Drolsbach and Pr{\"o}llochs(2023{\natexlab{a}})}]{Drolsbach.2023b}
Drolsbach, C.; and Pr{\"o}llochs, N. 2023{\natexlab{a}}.
\newblock Diffusion of Community Fact-Checked Misinformation on {T}witter.
\newblock In \emph{CSCW}.

\bibitem[{Drolsbach and Pr{\"o}llochs(2023{\natexlab{b}})}]{Drolsbach.2023}
Drolsbach, C.~P.; and Pr{\"o}llochs, N. 2023{\natexlab{b}}.
\newblock Believability and harmfulness shape the virality of misleading social
  media posts.
\newblock In \emph{WWW}.

\bibitem[{Drolsbach and Pr{\"o}llochs(2024)}]{Drolsbach.2024b}
Drolsbach, C.~P.; and Pr{\"o}llochs, N. 2024.
\newblock Content moderation on social media in the EU: insights from the DSA
  transparency database.
\newblock In \emph{WWW}.

\bibitem[{Drolsbach, Solovev, and Pr{\"o}llochs(2024)}]{Drolsbach.2024}
Drolsbach, C.~P.; Solovev, K.; and Pr{\"o}llochs, N. 2024.
\newblock Community notes increase trust in fact-checking on social media.
\newblock \emph{PNAS Nexus}, pgae217.

\bibitem[{Epstein, Pennycook, and Rand(2020)}]{Epstein.2020}
Epstein, Z.; Pennycook, G.; and Rand, D. 2020.
\newblock Will the crowd game the algorithm? {U}sing layperson judgments to
  combat misinformation on social media by downranking distrusted sources.
\newblock In \emph{CHI}.

\bibitem[{Feuerriegel et~al.(2023)Feuerriegel, DiResta, Goldstein, Kumar,
  Lorenz-Spreen, Tomz, and Pr{\"o}llochs}]{Feuerriegel.2023}
Feuerriegel, S.; DiResta, R.; Goldstein, J.~A.; Kumar, S.; Lorenz-Spreen, P.;
  Tomz, M.; and Pr{\"o}llochs, N. 2023.
\newblock Research can help to tackle AI-generated disinformation.
\newblock \emph{Nature Human Behaviour}, 7(11): 1818--1821.

\bibitem[{Feuerriegel et~al.(2025)Feuerriegel, Maarouf, B{\"a}r, Geissler,
  Schweisthal, Pr{\"o}llochs, Robertson, Rathje, Hartmann, Mohammad
  et~al.}]{Feuerriegel.2025}
Feuerriegel, S.; Maarouf, A.; B{\"a}r, D.; Geissler, D.; Schweisthal, J.;
  Pr{\"o}llochs, N.; Robertson, C.~E.; Rathje, S.; Hartmann, J.; Mohammad,
  S.~M.; et~al. 2025.
\newblock Using natural language processing to analyse text data in behavioural
  science.
\newblock \emph{Nature Reviews Psychology}, 4: 96--111.

\bibitem[{Frey and van~de Rijt(2021)}]{Frey.2021}
Frey, V.; and van~de Rijt, A. 2021.
\newblock Social influence undermines the wisdom of the crowd in sequential
  decision making.
\newblock \emph{Management Science}, 67(7): 4273--4286.

\bibitem[{Gallotti et~al.(2020)Gallotti, Valle, Castaldo, Sacco, and
  De~Domenico}]{Gallotti.2020}
Gallotti, R.; Valle, F.; Castaldo, N.; Sacco, P.; and De~Domenico, M. 2020.
\newblock Assessing the risks of 'infodemics' in response to {COVID-19}
  epidemics.
\newblock \emph{Nature Human Behaviour}, 4(12): 1285--1293.

\bibitem[{Geeng, Yee, and Roesner(2020)}]{Geeng.2020}
Geeng, C.; Yee, S.; and Roesner, F. 2020.
\newblock Fake News on {F}acebook and {T}witter: Investigating How People
  (Don't) Investigate.
\newblock In \emph{CHI}.

\bibitem[{Goodman-Bacon(2021)}]{GoodmanBacon.2021}
Goodman-Bacon, A. 2021.
\newblock Difference-in-differences with variation in treatment timing.
\newblock \emph{Journal of Econometrics}, 225(2): 254--277.

\bibitem[{Hutto, Yardi, and Gilbert(2013)}]{Hutto.2013}
Hutto, C.~J.; Yardi, S.; and Gilbert, E. 2013.
\newblock {A longitudinal study of follow predictors on twitter}.
\newblock In \emph{CHI}.

\bibitem[{Kaiser, Vaccari, and Chadwick(2022)}]{Kaiser.2022}
Kaiser, J.; Vaccari, C.; and Chadwick, A. 2022.
\newblock {Partisan Blocking: Biased Responses to Shared Misinformation
  Contribute to Network Polarization on Social Media}.
\newblock \emph{Journal of Commununication}, 72(2): 214--240.

\bibitem[{Kaushal et~al.(2024)Kaushal, Van De~Kerkhof, Goanta, Spanakis, and
  Iamnitchi}]{Kaushal.2024}
Kaushal, R.; Van De~Kerkhof, J.; Goanta, C.; Spanakis, G.; and Iamnitchi, A.
  2024.
\newblock Automated Transparency: A Legal and Empirical Analysis of the Digital
  Services Act Transparency Database.
\newblock In \emph{FAccT}.

\bibitem[{Kim and Dennis(2019)}]{Kim.2019}
Kim, A.; and Dennis, A.~R. 2019.
\newblock Says who? {T}he effects of presentation format and source rating on
  fake news in social media.
\newblock \emph{MIS Quarterly}, 43(3): 1025--1039.

\bibitem[{Kim et~al.(2025)Kim, Wang, Shi, Ling, and Evans}]{Kim.2025}
Kim, J.; Wang, Z.; Shi, H.; Ling, H.-K.; and Evans, J. 2025.
\newblock {Differential impact from individual versus collective misinformation
  tagging on the diversity of Twitter (X) information engagement and mobility}.
\newblock \emph{Nature Commununications}, 16(973): 1--14.

\bibitem[{Kivran-Swaine, Govindan, and Naaman(2011)}]{Kivran-Swaine.2011}
Kivran-Swaine, F.; Govindan, P.; and Naaman, M. 2011.
\newblock {The impact of network structure on breaking ties in online social
  networks: unfollowing on twitter}.
\newblock In \emph{CHI}.

\bibitem[{Kwak, Chun, and Moon(2011)}]{Kwak.2011}
Kwak, H.; Chun, H.; and Moon, S. 2011.
\newblock {Fragile online relationship: a first look at unfollow dynamics in
  twitter}.
\newblock In \emph{CHI}.

\bibitem[{Lazer et~al.(2018)Lazer, Baum, Benkler, Berinsky, Greenhill, Menczer,
  Metzger, Nyhan, Pennycook, Rothschild, Schudson, Sloman, Sunstein, Thorson,
  Watts, and Zittrain}]{Lazer.2018}
Lazer, D. M.~J.; Baum, M.~A.; Benkler, Y.; Berinsky, A.~J.; Greenhill, K.~M.;
  Menczer, F.; Metzger, M.~J.; Nyhan, B.; Pennycook, G.; Rothschild, D.;
  Schudson, M.; Sloman, S.~A.; Sunstein, C.~R.; Thorson, E.~A.; Watts, D.~J.;
  and Zittrain, J.~L. 2018.
\newblock The science of fake news.
\newblock \emph{Science}, 359(6380): 1094--1096.

\bibitem[{Loureiro et~al.(2022)Loureiro, Barbieri, Neves, Anke, and
  Camacho-Collados}]{Loureiro.2022}
Loureiro, D.; Barbieri, F.; Neves, L.; Anke, L.~E.; and Camacho-Collados, J.
  2022.
\newblock {TimeLMs}: Diachronic Language Models from {T}witter.
\newblock \emph{arXiv}.

\bibitem[{Ma et~al.(2016)Ma, Gao, Mitra, Kwon, Jansen, Wong, and Cha}]{Ma.2016}
Ma, J.; Gao, W.; Mitra, P.; Kwon, S.; Jansen, B.~J.; Wong, K.-F.; and Cha, M.
  2016.
\newblock Detecting rumors from microblogs with recurrent neural networks.
\newblock In \emph{ICJAI}.

\bibitem[{McCabe et~al.(2024)McCabe, Ferrari, Green, Lazer, and
  Esterling}]{McCabe.2024}
McCabe, S.~D.; Ferrari, D.; Green, J.; Lazer, D. M.~J.; and Esterling, K.~M.
  2024.
\newblock {Post-January 6th deplatforming reduced the reach of misinformation
  on Twitter}.
\newblock \emph{Nature}, 630: 132--140.

\bibitem[{McPherson, Smith-Lovin, and Cook(2001)}]{McPherson.2001}
McPherson, M.; Smith-Lovin, L.; and Cook, J.~M. 2001.
\newblock Birds of a feather: Homophily in social networks.
\newblock \emph{Annual Review of Sociology}, 27: 415--444.

\bibitem[{Meta(2025)}]{Meta.2025}
Meta. 2025.
\newblock {Testing Begins for Community Notes on Facebook, Instagram and
  Threads}.
\newblock
  https://about.fb.com/news/2025/03/testing-begins-community-notes-facebook-instagram-threads.

\bibitem[{Micallef et~al.(2022)Micallef, Armacost, Memon, and
  Patil}]{Micallef.2022}
Micallef, N.; Armacost, V.; Memon, N.; and Patil, S. 2022.
\newblock True or false: Studying the work practices of professional
  fact-checkers.
\newblock In \emph{CSCW}.

\bibitem[{Moravec, Minas, and Dennis(2019)}]{Moravec.2019}
Moravec, P.~L.; Minas, R.~K.; and Dennis, A. 2019.
\newblock Fake news on social media: People believe what they want to believe
  when it makes no sense at all.
\newblock \emph{MIS Quarterly}, 43(4): 1343--1360.

\bibitem[{Nyhan et~al.(2020)Nyhan, Porter, Reifler, and Wood}]{Nyhan.2020}
Nyhan, B.; Porter, E.; Reifler, J.; and Wood, T.~J. 2020.
\newblock {Taking Fact-Checks Literally But Not Seriously? The Effects of
  Journalistic Fact-Checking on Factual Beliefs and Candidate Favorability}.
\newblock \emph{Political Behavior}, 42(3): 939--960.

\bibitem[{Oh, Agrawal, and Rao(2013)}]{Oh.2013}
Oh, O.; Agrawal, M.; and Rao, H.~R. 2013.
\newblock Community intelligence and social media services: A rumor theoretic
  analysis of tweets during social crises.
\newblock \emph{MIS Quarterly}, 37(2): 407--426.

\bibitem[{Pennycook, Cannon, and Rand(2018)}]{Pennycook.2018}
Pennycook, G.; Cannon, T.~D.; and Rand, D.~G. 2018.
\newblock Prior exposure increases perceived accuracy of fake news.
\newblock \emph{Journal of Experimental Psychology: General}, 147(12):
  1865--1880.

\bibitem[{Pennycook et~al.(2021)Pennycook, Epstein, Mosleh, Arechar, Eckles,
  and Rand}]{Pennycook.2021}
Pennycook, G.; Epstein, Z.; Mosleh, M.; Arechar, A.~A.; Eckles, D.; and Rand,
  D.~G. 2021.
\newblock Shifting attention to accuracy can reduce misinformation online.
\newblock \emph{Nature}, 592(7855): 590--595.

\bibitem[{Pennycook et~al.(2020)Pennycook, McPhetres, Zhang, Lu, and
  Rand}]{Pennycook.2020b}
Pennycook, G.; McPhetres, J.; Zhang, Y.; Lu, J.~G.; and Rand, D.~G. 2020.
\newblock Fighting {COVID-19} misinformation on social media: Experimental
  evidence for a scalable accuracy-nudge intervention.
\newblock \emph{Psychological Science}, 31(7): 770--780.

\bibitem[{Pennycook and Rand(2019{\natexlab{a}})}]{Pennycook.2019}
Pennycook, G.; and Rand, D.~G. 2019{\natexlab{a}}.
\newblock Fighting misinformation on social media using crowdsourced judgments
  of news source quality.
\newblock \emph{PNAS}, 116(7): 2521--2526.

\bibitem[{Pennycook and Rand(2019{\natexlab{b}})}]{Pennycook.2019b}
Pennycook, G.; and Rand, D.~G. 2019{\natexlab{b}}.
\newblock Lazy, not biased: Susceptibility to partisan fake news is better
  explained by lack of reasoning than by motivated reasoning.
\newblock \emph{Cognition}, 188: 39--50.

\bibitem[{{Pew Research Center}(2024{\natexlab{a}})}]{Pew.2024a}
{Pew Research Center}. 2024{\natexlab{a}}.
\newblock Many Americans find value in getting news on social media, but
  concerns about inaccuracy have risen.
\newblock
  https://www.pewresearch.org/short-reads/2024/02/07/many-americans-find-value-in-getting-news-on-social-media-but-concerns-about-inaccuracy-have-risen/.

\bibitem[{{Pew Research Center}(2024{\natexlab{b}})}]{Pew.2024b}
{Pew Research Center}. 2024{\natexlab{b}}.
\newblock Social media and news fact sheet.
\newblock
  https://www.pewresearch.org/journalism/fact-sheet/social-media-and-news-fact-sheet/.

\bibitem[{Pfeffer, Matter, and Sargsyan(2023)}]{Pfeffer.2023}
Pfeffer, J.; Matter, D.; and Sargsyan, A. 2023.
\newblock The half-life of a tweet.
\newblock In \emph{ICWSM}.

\bibitem[{Pilarski, Solovev, and Pr{\"o}llochs(2024)}]{Pilarski.2024}
Pilarski, M.; Solovev, K.; and Pr{\"o}llochs, N. 2024.
\newblock Community notes vs. snoping: How the crowd selects fact-checking
  targets on social media.
\newblock In \emph{ICWSM}.

\bibitem[{Poynter(2019)}]{Poynter.2019}
Poynter. 2019.
\newblock Most {Republicans} don't trust fact-checkers, and most {Americans}
  don’t trust the media.
\newblock
  \url{https://www.poynter.org/ifcn/2019/most-republicans-dont-trust-fact-checkers-and-most-americans-dont-trust-the-media/}.

\bibitem[{Pr{\"o}llochs(2022)}]{Prollochs.2022a}
Pr{\"o}llochs, N. 2022.
\newblock Community-based fact-checking on {T}witter's {B}irdwatch platform.
\newblock In \emph{ICWSM}.

\bibitem[{Pr{\"o}llochs, B{\"a}r, and Feuerriegel(2021)}]{Prollochs.2021a}
Pr{\"o}llochs, N.; B{\"a}r, D.; and Feuerriegel, S. 2021.
\newblock Emotions in online rumor diffusion.
\newblock \emph{EPJ Data Science}, 10(1).
\newblock 51.

\bibitem[{Pr{\"o}llochs and Feuerriegel(2023)}]{Prollochs.2022b}
Pr{\"o}llochs, N.; and Feuerriegel, S. 2023.
\newblock Mechanisms of true and false rumor sharing in social media:
  {C}ollective intelligence or herd behavior?
\newblock In \emph{CSCW}.

\bibitem[{Resnick et~al.(2021)Resnick, Alfayez, Im, and Gilbert}]{Resnick.2021}
Resnick, P.; Alfayez, A.; Im, J.; and Gilbert, E. 2021.
\newblock Informed crowds can effectively identify misinformation.
\newblock \emph{arXiv}, (2108.07898).

\bibitem[{Rivers and Lewis(2014)}]{Rivers.2014}
Rivers, C.~M.; and Lewis, B.~L. 2014.
\newblock Ethical research standards in a world of big data.
\newblock \emph{F1000Research}, 3: 38.

\bibitem[{Shao et~al.(2016)Shao, Ciampaglia, Flammini, and Menczer}]{Shao.2016}
Shao, C.; Ciampaglia, G.~L.; Flammini, A.; and Menczer, F. 2016.
\newblock Hoaxy: A platform for tracking online misinformation.
\newblock In \emph{WWW Companion}.

\bibitem[{Shore, Baek, and Dellarocas(2018)}]{Shore.2018}
Shore, J.; Baek, J.; and Dellarocas, C. 2018.
\newblock Network structure and patterns of information diversity on {T}witter.
\newblock \emph{MIS Quarterly}, 42(3): 849--972.

\bibitem[{Slaughter et~al.(2025)Slaughter, Peytavin, Ugander, and
  Saveski}]{Slaughter2025}
Slaughter, I.; Peytavin, A.; Ugander, J.; and Saveski, M. 2025.
\newblock Community notes moderate engagement with and diffusion of false
  information online.
\newblock \emph{arXiv}.

\bibitem[{Solovev and Pr{\"o}llochs(2022)}]{Solovev.2022b}
Solovev, K.; and Pr{\"o}llochs, N. 2022.
\newblock Moral emotions shape the virality of COVID-19 misinformation on
  social media.
\newblock In \emph{WWW}.

\bibitem[{Solovev and Pr{\"o}llochs(2025)}]{Solovev.2025}
Solovev, K.; and Pr{\"o}llochs, N. 2025.
\newblock References to unbiased sources increase the helpfulness of community
  fact-checks.
\newblock \emph{arXiv}.

\bibitem[{Starbird et~al.(2014)Starbird, Maddock, Orand, Achterman, and
  Mason}]{Starbird.2014}
Starbird, K.; Maddock, J.; Orand, M.; Achterman, P.; and Mason, R.~M. 2014.
\newblock Rumors, false flags, and digital vigilantes: Misinformation on
  {T}witter after the 2013 {B}oston marathon bombing.
\newblock In \emph{iConference}.

\bibitem[{Sun and Abraham(2021)}]{Sun.2021}
Sun, L.; and Abraham, S. 2021.
\newblock Estimating dynamic treatment effects in event studies with
  heterogeneous treatment effects.
\newblock \emph{Journal of Econometrics}, 225(2): 175--199.

\bibitem[{Swire-Thompson et~al.(2020)Swire-Thompson, Ecker, Lewandowsky, and
  Berinsky}]{Swire-Thompson.2020}
Swire-Thompson, B.; Ecker, U. K.~H.; Lewandowsky, S.; and Berinsky, A.~J. 2020.
\newblock {They Might Be a Liar But They{'}re My Liar: Source Evaluation and
  the Prevalence of Misinformation}.
\newblock \emph{Political Psychology}, 41(1): 21--34.

\bibitem[{TikTok(2025)}]{TikTok.2025}
TikTok. 2025.
\newblock {Testing a new feature to enhance content on TikTok - Newsroom
  {$\vert$} TikTok}.
\newblock https://newsroom.tiktok.com/en-us/footnotes.

\bibitem[{Trujillo, Fagni, and Cresci(2025)}]{Trujillo.2025}
Trujillo, A.; Fagni, T.; and Cresci, S. 2025.
\newblock The DSA Transparency Database: Auditing Self-reported Moderation
  Actions by Social Media.
\newblock In \emph{CSCW}.

\bibitem[{Van~Bavel et~al.(2024)Van~Bavel, Robertson, del Rosario, Rasmussen,
  and Rathje}]{VanBavel.2024}
Van~Bavel, J.~J.; Robertson, C.~E.; del Rosario, K.; Rasmussen, J.; and Rathje,
  S. 2024.
\newblock {Social Media and Morality}.
\newblock \emph{Annual Review of Psychology}, (Volume 75, 2024): 311--340.

\bibitem[{Vo and Lee(2018)}]{Vo.2018}
Vo, N.; and Lee, K. 2018.
\newblock The rise of guardians: Fact-checking url recommendation to combat
  fake news.
\newblock In \emph{SIGIR}.

\bibitem[{Vosoughi, Roy, and Aral(2018)}]{Vosoughi.2018}
Vosoughi, S.; Roy, D.; and Aral, S. 2018.
\newblock The spread of true and false news online.
\newblock \emph{Science}, 359(6380): 1146--1151.

\bibitem[{Wojcik et~al.(2022)Wojcik, Hilgard, Judd, Mocanu, Ragain, Hunzaker,
  Coleman, and Baxter}]{Wojcik.2022}
Wojcik, S.; Hilgard, S.; Judd, N.; Mocanu, D.; Ragain, S.; Hunzaker, M.;
  Coleman, K.; and Baxter, J. 2022.
\newblock Birdwatch: Crowd wisdom and bridging algorithms can inform
  understanding and reduce the spread of misinformation.
\newblock \emph{arXiv}.

\bibitem[{{World Economic Forum}(2024)}]{WEF.2024}
{World Economic Forum}. 2024.
\newblock Global risks report.
\newblock \url{https://www.weforum.org/publications/global-risks-report-2024/}.

\bibitem[{Wu et~al.(2019)Wu, Morstatter, Carley, and Liu}]{Wu.2019}
Wu, L.; Morstatter, F.; Carley, K.~M.; and Liu, H. 2019.
\newblock Misinformation in social media: definition, manipulation, and
  detection.
\newblock \emph{SIGKDD Explorations Newsletter}, 21(2): 80--90.

\bibitem[{X(2021)}]{Twitter.2021}
X. 2021.
\newblock Introducing {B}irdwatch, a Community-Based Approach to
  Misinformation.
\newblock
  \url{https://blog.twitter.com/en_us/topics/product/2021/introducing-birdwatch-a-community-based-approach-to-misinformation}.

\bibitem[{X(2024)}]{CNsFAQ.2024}
X. 2024.
\newblock FAQs for {Community Notes} program.
\newblock \url{https://communitynotes.x.com/guide/en/about/faq}.

\bibitem[{Xu et~al.(2013)Xu, Huang, Kwak, and Contractor}]{Xu.2013}
Xu, B.; Huang, Y.; Kwak, H.; and Contractor, N. 2013.
\newblock {Structures of broken ties: exploring unfollow behavior on twitter}.
\newblock In \emph{CSCW}.

\bibitem[{YouTube(2024)}]{Youtube.2024}
YouTube. 2024.
\newblock {Testing new ways to offer viewers more context and information on
  videos}.
\newblock
  https://blog.youtube/news-and-events/new-ways-to-offer-viewers-more-context.

\end{thebibliography}

\clearpage
\onecolumn
\begin{center}
\LARGE
\textbf{Supplementary Materials}
\end{center}
\normalsize
\renewcommand\thetable{S\arabic{table}}
\setcounter{table}{0}
\renewcommand\thefigure{S\arabic{figure}}
\setcounter{figure}{0}

\section{Dataset Overview}
\label{sec:data_overview}

An overview of the dataset used in this study is shown in Table \ref{tab:data_summary}.

\begin{table}[h!]
\centering

\begin{tabularx}{0.85\textwidth}{@{\extracolsep{\fill}}lccc}
\toprule
 & (1) & (2) & (3) \\
 & All & Treated & Never-Treated \\
\midrule

\#Posts & \num{3516} & \num{2671} & \num{845} \\
\#Users & \num{2142} & \num{1782} & \num{670} \\
Post Date & {02/27/2024 -- 08/29/2024} & {02/27/2024 -- 08/29/2024} & {02/29/2024 -- 08/29/2024} \\
Note Date & {03/01/2024 -- 08/29/2024} & {03/01/2024 -- 08/29/2024} & {03/02/2024 -- 08/29/2024} \\[2.5mm]

\textit{User Characteristics} & & & \\[-2.5mm]
\rule{3cm}{0.1pt} \\

Account Age (Years) & \num{9.61} (\num{5.13}) & \num{9.48} (\num{5.11}) & \num{10.03} (\num{5.16}) \\
Verified & \SI{13}{\percent} & \SI{13}{\percent} & \SI{14}{\percent} \\
\#Tweets & \num{74246.70} (\num{121494.61}) & \num{74273.48} (\num{122090.30}) & \num{74162.06} (\num{119663.64}) \\
\#Followers & \num{843065.58} (\num{2680902.74}) & \num{799273.34} (\num{2700249.07}) & \num{981490.53} (\num{2615561.75}) \\
\#Followees & \num{6075.49} (\num{27191.04}) & \num{5358.13} (\num{19661.35}) & \num{8343.01} (\num{43005.66}) \\[2.5mm]

\textit{Post Characteristics} & & & \\[-2.5mm]
\rule{3cm}{0.1pt} \\

\#Words & \num{25.75} (\num{15.05}) & \num{25.01} (\num{14.96}) & \num{28.08} (\num{15.12}) \\
\#URLs & \num{1.01} (\num{0.54}) & \num{1.01} (\num{0.52}) & \num{1.01} (\num{0.62}) \\
\#Reposts & \num{1420.82} (\num{2849.51}) & \num{1289.08} (\num{2546.79}) & \num{1837.24} (\num{3615.05}) \\
\#Replies & \num{913.67} (\num{2280.41}) & \num{918.03} (\num{2397.06}) & \num{899.91} (\num{1865.49}) \\
\#Likes & \num{9491.73} (\num{21607.12}) & \num{9737.39} (\num{22218.92}) & \num{8715.24} (\num{19539.81}) \\
\#Quotes & \num{334.81} (\num{786.02}) & \num{357.15} (\num{709.35}) & \num{264.21} (\num{987.26}) \\
Media & \SI{67}{\percent} & \SI{69}{\percent} & \SI{61}{\percent} \\
Sentiment & \SI{27}{\percent} & \SI{29}{\percent} & \SI{23}{\percent} \\
Political & \SI{21}{\percent} & \SI{19}{\percent} & \SI{27}{\percent} \\
\bottomrule
\end{tabularx}
\caption{\textbf{Descriptive Statistics of User and Post Characteristics.} Summary statistics are displayed for (1) the overall dataset, (2) treated units, (3) and never-treated units. Binary features are reported as shares, while continuous features are described by their mean values (standard deviations in parentheses).}
\label{tab:data_summary}
\end{table}

\newpage
\section{Estimation Results}

Tables~\ref{tab:group_att_growth_rate} and~\ref{tab:event_time_att_growth_rate} present the estimated group-time and event-time average treatment effects using daily follower growth as the outcome. Tables~\ref{tab:group_att_follower_count} and~\ref{tab:event_time_att_follower_count} report the corresponding results using the log number of followers.

\begin{table}[h!]
\centering

\begin{tabular}{
  l
  S[table-format=1.2]
  S[table-format=1.2]
  S[table-format=1.2]
  S[table-format=1.2]
  S[table-format=1.2]
  S[table-format=1.2]
  S[table-format=1.2]
  S[table-format=1.2]
  S[table-format=1.2]
}
\toprule
\multicolumn{10}{l}{Dependent Variable: Daily Follower Growth Rate} \\
\midrule
& {Main} & \multicolumn{4}{c}{Robustness} & \multicolumn{4}{c}{Sensitivity} \\
\cmidrule(lr){2-2} \cmidrule(lr){3-6} \cmidrule(lr){7-10}
 & {(1)} & {(2)} & {(3)} & {(4)} & {(5)} & {(6)} & {(7)} & {(8)} & {(9)} \\
\midrule
$t + 0$ & -0.04 & -0.05 &  0.00 & 0.00 & -0.05 & -0.08{$^{*}$} & -0.03 & -0.02 & -0.05 \\
        & (0.05) & (0.05) & (0.02) & (0.07) & (0.06) & (0.03) & (0.09) & (0.09) & (0.06) \\
$t + 1$ &  0.02 &  0.00 &  0.02 &  0.00 &  0.02 & -0.05 &  0.16 &  0.20 & -0.04 \\
        & (0.12) & (0.06) & (0.02) & (0.12) & (0.06) & (0.13) & (0.42) & (0.22) & (0.06) \\
$t + 2$ &  0.05 & -0.07 & 0.00 &  0.00 &  0.05 &  0.04 &  0.21 &  0.08 &  0.03 \\
        & (0.20) & (0.19) & (0.03) & (0.27) & (0.21) & (0.05) & (0.36) & (0.13) & (0.23) \\
$t + 3$ &  0.09 &  0.00 & -0.06 &  0.08 &  0.09 &  0.02 &  0.20 &  0.12 &  0.06 \\
        & (0.19) & (0.18) & (0.19) & (0.20) & (0.18) & (0.05) & (0.27) & (0.14) & (0.20) \\
\addlinespace[4pt]
Average & 0.01 & -0.02 & 0.01 & 0.01 & 0.01 & -0.04 & 0.12 & 0.10 & -0.03 \\
                      & (0.07) & (0.05) & (0.02) & (0.09) & (0.07) & (0.07) & (0.27) & (0.12) & (0.05) \\
\midrule
Covariates           & {Yes} & {No} & {Yes} & {Yes} & {Yes} & {Yes} & {Yes} & {Yes} & {Yes} \\
Control group        & {Not-yet} & {Not-yet} & {Not-yet} & {Not-yet} & {Never} & {Not-yet} & {Not-yet} & {Not-yet} & {Not-yet} \\
Sub-sample & {No} & {No} & {Top \SI{10}{\percent}} & {Treated} & {No} &{Large} & {Small} & {Pol.}
& {Non-pol.} \\
 &  &  & {removed} & {once} &  & {accounts} & {accounts} & {posts}
& {posts} \\

\midrule 
\#Observations       & \num{73836} & \num{73836} & \num{65667} &  \num{39165} & \num{73836} & \num{37170} & \num{36666} & \num{15435} & \num{58401} \\
\#Posts              & \num{3516} & \num{3516} & \num{3127} & \num{1865} & \num{3516} & \num{1770} & \num{1,746} & \num{735} & \num{2781} \\
\bottomrule
\multicolumn{10}{l}{\footnotesize{$^{*}p<0.05$}}
\end{tabular}
\caption{\textbf{Group ATTs for Daily Growth Rate.} Displayed are group-specific ATTs, along with the overall average ATT, estimated using staggered difference-in-differences models with daily follower growth as the outcome variable. Group-specific ATTs refer to treatment effects for posts first treated on day \num{0}, \num{1}, \num{2}, and \num{3} after post publication. Each column presents a different model specification, including sensitivity analyses by follower size and political content. Clustered (bootstrapped) standard errors are reported in parentheses.}
\label{tab:group_att_growth_rate}
\end{table}

\begin{table}
\centering

\begin{tabular}{
  l
  S[table-format=1.2]
  S[table-format=1.2]
  S[table-format=1.2]
  S[table-format=1.2]
  S[table-format=1.2]
  S[table-format=1.2]
  S[table-format=1.2]
  S[table-format=1.2]
  S[table-format=1.2]
}
\toprule
\multicolumn{10}{l}{Dependent Variable: Daily Follower Growth Rate} \\
\midrule
& {Main} & \multicolumn{4}{c}{Robustness} & \multicolumn{4}{c}{Sensitivity} \\
\cmidrule(lr){2-2} \cmidrule(lr){3-6} \cmidrule(lr){7-10}
            & {(1)} & {(2)} & {(3)} & {(4)} & {(5)} & {(6)} & {(7)} & {(8)} & {(9)} \\
 \midrule
Day$_{-7}$ & 0.01 & 0.01 & 0.03 & 0.01 & 0.00 & 0.03 & -0.02 & 0.00 & 0.01 \\
           & (0.03) & (0.03) & (0.02) & (0.04) & (0.04) & (0.05) & (0.03) & (0.07) & (0.03) \\
Day$_{-6}$ & 0.00 & 0.01 & 0.00 & 0.02 & -0.01 & 0.00 & 0.02 & 0.03 & 0.00 \\
           & (0.03) & (0.02) & (0.01) & (0.04) & (0.05) & (0.01) & (0.04) & (0.04) & (0.03) \\
Day$_{-5}$ & -0.01 & -0.01 & -0.01 & -0.01 & -0.04 & -0.01 & -0.02 & -0.02 & -0.01 \\
           & (0.03) & (0.03) & (0.01) & (0.04) & (0.05) & (0.01) & (0.05) & (0.04) & (0.04) \\
Day$_{-4}$ & -0.03 & -0.03 & 0.00 & -0.05 & -0.10 & 0.00 & -0.07 & 0.02 & -0.05 \\
           & (0.04) & (0.03) & (0.01) & (0.05) & (0.19) & (0.01) & (0.08) & (0.04) & (0.05) \\
Day$_{-3}$ & 0.03 & 0.03 & 0.02 & 0.04 & 0.04 & 0.02 & 0.04 & 0.00 & 0.04 \\
           & (0.02) & (0.02) & (0.01) & (0.02) & (0.04) & (0.02) & (0.03) & (0.03) & (0.02) \\
Day$_{-2}$ & 0.01 & 0.01 & -0.02 & -0.02 & 0.09 & 0.00 & 0.03 & -0.02 & 0.02 \\
           & (0.04) & (0.03) & (0.01) & (0.05) & (0.20) & (0.02) & (0.07) & (0.03) & (0.05) \\
Day$_{-1}$ & -0.01 & 0.00 & 0.01 & 0.05 & 0.01 & 0.02 & -0.06 & -0.03 & -0.01 \\
           & (0.05) & (0.04) & (0.01) & (0.05) & (0.11) & (0.07) & (0.07) & (0.10) & (0.05) \\
Day$_{0}$ & 0.07 & 0.11 & 0.04 & 0.00 & 0.06 & 0.05 & 0.08 & 0.06 & 0.07 \\
          & (0.07) & (0.06) & (0.02) & (0.10) & (0.08) & (0.07) & (0.11) & (0.05) & (0.08) \\
Day$_{1}$ & 0.02 & 0.04 & 0.02 & -0.02 & 0.00 & 0.01 & 0.00 & 0.07 & -0.01 \\
          & (0.07) & (0.06) & (0.03) & (0.10) & (0.10) & (0.03) & (0.16) & (0.05) & (0.09) \\
Day$_{2}$ & 0.01 & -0.02 & -0.01 & -0.01 & 0.00 & -0.03 & 0.09 & 0.12 & -0.03 \\
          & (0.07) & (0.05) & (0.02) & (0.10) & (0.08) & (0.08) & (0.28) & (0.11) & (0.06) \\
Day$_{3}$ & -0.01 & -0.05 & -0.01 & 0.00 & -0.01 & -0.06 & 0.09 & 0.12 & -0.06 \\
          & (0.08) & (0.05) & (0.02) & (0.09) & (0.08) & (0.08) & (0.29) & (0.13) & (0.06) \\
Day$_{4}$ & -0.02 & -0.04 & -0.01 & -0.02 & -0.02 & -0.09 & 0.10 & 0.10 & -0.07 \\
          & (0.08) & (0.05) & (0.02) & (0.10) & (0.09) & (0.07) & (0.30) & (0.14) & (0.07) \\
Day$_{5}$ & -0.03 & -0.05 & 0.00 & -0.02 & -0.03 & -0.10 & 0.11 & 0.07 & -0.08 \\
          & (0.09) & (0.06) & (0.02) & (0.11) & (0.09) & (0.10) & (0.30) & (0.14) & (0.08) \\
Day$_{6}$ & 0.00 & -0.05 & 0.01 & 0.05 & 0.00 & -0.08 & 0.13 & 0.08 & -0.05 \\
          & (0.08) & (0.05) & (0.02) & (0.10) & (0.09) & (0.08) & (0.30) & (0.14) & (0.06) \\
Day$_{7}$ & 0.03 & -0.03 & 0.01 & 0.03 & 0.03 & -0.05 & 0.16 & 0.13 & -0.01 \\
          & (0.08) & (0.06) & (0.02) & (0.10) & (0.08) & (0.08) & (0.29) & (0.14) & (0.07) \\
Day$_{8}$ & -0.01 & -0.06 & 0.00 & 0.00 & -0.01 & -0.06 & 0.11 & 0.14 & -0.05 \\
          & (0.08) & (0.06) & (0.02) & (0.10) & (0.09) & (0.08) & (0.30) & (0.14) & (0.06) \\
Day$_{9}$ & 0.02 & -0.04 & 0.01 & 0.01 & 0.02 & -0.05 & 0.15 & 0.13 & -0.02 \\
          & (0.08) & (0.06) & (0.02) & (0.10) & (0.09) & (0.08) & (0.30) & (0.15) & (0.06) \\
Day$_{10}$ & 0.05 & -0.01 & 0.02 & 0.04 & 0.05 & -0.04 & 0.20 & 0.13 & 0.02 \\
           & (0.08) & (0.05) & (0.02) & (0.10) & (0.09) & (0.08) & (0.30) & (0.15) & (0.06) \\
Day$_{11}$ & 0.03 & -0.03 & 0.00 & 0.02 & 0.03 & -0.04 & 0.16 & 0.08 & 0.00 \\
           & (0.08) & (0.05) & (0.02) & (0.10) & (0.09) & (0.08) & (0.30) & (0.15) & (0.06) \\
Day$_{12}$ & 0.01 & -0.04 & 0.00 & 0.01 & 0.01 & -0.04 & 0.12 & 0.08 & -0.02 \\
           & (0.08) & (0.05) & (0.02) & (0.11) & (0.09) & (0.09) & (0.32) & (0.15) & (0.06) \\
Day$_{13}$ & 0.00 & -0.04 & 0.00 & 0.02 & 0.00 & -0.07 & 0.10 & 0.11 & -0.04 \\
           & (0.08) & (0.05) & (0.02) & (0.09) & (0.09) & (0.10) & (0.32) & (0.15) & (0.06) \\
Day$_{14}$ & -0.06 & -0.09 & -0.01 & -0.02 & -0.06 & -0.07 & -0.08 & 0.02 & -0.09 \\
           & (0.06) & (0.05) & (0.02) & (0.08) & (0.06) & (0.04) & (0.11) & (0.10) & (0.08) \\

\midrule
Covariates           & {Yes} & {No} & {Yes} & {Yes} & {Yes} & {Yes} & {Yes} & {Yes} & {Yes} \\
Control group        & {Not-yet} & {Not-yet} & {Not-yet} & {Not-yet} & {Never} & {Not-yet} & {Not-yet} & {Not-yet} & {Not-yet} \\
Sub-sample & {No} & {No} & {Top \SI{10}{\percent}} & {Treated} & {No} &{Large} & {Small} & {Pol.}
& {Non-pol.} \\
 &  &  & {removed} & {once} &  &{accounts} & {accounts} & {posts}
& {posts} \\

\midrule 
\#Observations       & \num{73836} & \num{73836} & \num{65667} &  \num{39165} & \num{73836} & \num{37170} & \num{36666} & \num{15435} & \num{58401} \\
\#Posts              & \num{3516} & \num{3516} & \num{3127} & \num{1865} & \num{3516} & \num{1770} & \num{1,746} & \num{735} & \num{2781} \\
\bottomrule
\multicolumn{10}{l}{\footnotesize{$^{*}p<0.05$}}
\end{tabular}
\caption{\textbf{Event-study ATTs for Daily Growth Rate.} The table displays event-study ATTs for the daily follower growth rate as outcome variable, from seven days before to fourteen days after receiving a community note. Each column presents a different model specification, including sensitivity analyses by follower size and political content. Clustered (bootstrapped) standard errors are reported in parentheses.}
\label{tab:event_time_att_growth_rate}
\end{table}

\begin{table}
\centering
\begin{tabular}{
  l
  S[table-format=1.4]
  S[table-format=1.4]
  S[table-format=1.4]
  S[table-format=1.4]
  S[table-format=1.4]
  S[table-format=1.4]
  S[table-format=1.4]
  S[table-format=1.4]
  S[table-format=1.4]
}
\toprule
\multicolumn{10}{l}{Dependent Variable: Number of Followers (Log)} \\
\midrule
& {Main} & \multicolumn{4}{c}{Robustness} & \multicolumn{4}{c}{Sensitivity} \\
\cmidrule(lr){2-2} \cmidrule(lr){3-6} \cmidrule(lr){7-10}
& {(1)} & {(2)} & {(3)} & {(4)} & {(5)} & {(6)} & {(7)} & {(8)} & {(9)} \\
\midrule
$t + 0$ & -0.0026 & 0.0004 & 0.0003 & -0.0059 & -0.0028 & -0.0038 & -0.0059 & -0.0046 & -0.0025 \\
        & (0.0035) & (0.0024) & (0.0009) & (0.0041) & (0.0037) & (0.0024) & (0.0097) & (0.0067) & (0.0040) \\
$t + 1$ & -0.0007 & 0.0037 & 0.0009 & -0.0039 & -0.0008 & 0.0006 & -0.0065 & 0.0011 & -0.0015 \\
        & (0.0037) & (0.0025) & (0.0010) & (0.0041) & (0.0038) & (0.0046) & (0.0098) & (0.0039) & (0.0051) \\
$t + 2$ & -0.0013 & 0.0007 & 0.0002 & -0.0025 & -0.0014 & -0.0018 & -0.0035 & -0.0018 & -0.0016 \\
        & (0.0027) & (0.0023) & (0.0009) & (0.0034) & (0.0027) &(0.0028) & (0.0049) & (0.0028) & (0.0034) \\
$t + 3$ & -0.0054 & -0.0034 & 0.0001& -0.0066{$^{*}$} & -0.0054 & -0.0047{$^{*}$} & -0.0060 & -0.0015 & -0.0060 \\
        & (0.0025) & (0.0021) & (0.0013) & (0.0023) & (0.0024) & (0.0021) & (0.0033) & (0.0037) & (0.0027) \\
\addlinespace[4pt]
Average & -0.0016 & 0.0020 & 0.0006 & -0.0034 & -0.0017 & -0.0013 & -0.0059 & -0.0014 & -0.0020 \\
        & (0.0031) & (0.0019) & (0.0007) & (0.0033) & (0.0031) & (0.0027) & (0.0078) & (0.0039) & (0.0039) \\
\midrule
Covariates & {Yes} & {No} & {Yes} & {Yes} & {Yes} & {Yes} & {Yes} & {Yes} & {Yes} \\
Control group & {Not-yet} & {Not-yet} & {Not-yet} & {Not-yet} & {Never} & {Not-yet} & {Not-yet} & {Not-yet} & {Not-yet} \\

Sub-sample & {No} & {No} & {Top \SI{10}{\percent}} & {Treated} & {No} &{Large} & {Small} & {Pol.}
& {Non-pol.} \\
 &  &  & {removed} & {once} &  &{accounts} & {accounts} & {posts}
& {posts} \\ 

\midrule

\#Observations       & \num{73836} & \num{73836} & \num{65667} &  \num{39165} & \num{73836} & \num{37170} & \num{36666} & \num{15435} & \num{58401} \\
\#Posts              & \num{3516} & \num{3516} & \num{3127} & \num{1865} & \num{3516} & \num{1770} & \num{1,746} & \num{735} & \num{2781} \\
\bottomrule
\multicolumn{10}{l}{\footnotesize{$^{*}p<0.05$}}
\end{tabular}
\caption{\textbf{Group ATTs for Daily Follower Count (Log)}. Displayed are group-specific ATTs, along with the overall average ATT, estimated using staggered difference-in-differences models with daily follower growth as the outcome variable. Group-specific ATTs refer to treatment effects for posts first treated on day \num{0}, \num{1}, \num{2}, and \num{3} after post publication. Each column presents a different model specification, including sensitivity analyses by follower size and political content. Clustered (bootstrapped) standard errors are reported in parentheses.}
\label{tab:group_att_follower_count}
\end{table}

\begin{table}
\centering
\begin{tabular}{
  l
  S[table-format=1.4]
  S[table-format=1.4]
  S[table-format=1.4]
  S[table-format=1.4]
  S[table-format=1.4]
  S[table-format=1.4]
  S[table-format=1.4]
  S[table-format=1.4]
  S[table-format=1.4]
}
\toprule
\multicolumn{6}{l}{Dependent Variable: Number of Followers (Log)} \\
\midrule
& {Main} & \multicolumn{4}{c}{Robustness} & \multicolumn{4}{c}{Sensitivity} \\
\cmidrule(lr){2-2} \cmidrule(lr){3-6} \cmidrule(lr){7-10}
               & {(1)} & {(2)}  & {(3)}    & {(4)}  & {(5)}  & {(6)}  & {(7)}   & {(8)}    & {(9)} \\
\midrule
Day$_{-7}$  & -0.0005 & -0.0004 & 0.0000 & -0.0009 & -0.0005 & 0.0001 & -0.0010 & -0.0003 & -0.0005 \\
          & (0.0003) & (0.0004) & (0.0003) & (0.0004) & (0.0005) & (0.0003) & (0.0005) & (0.0006) & (0.0004) \\
Day$_{-6}$  &  0.0000 &  0.0000 &  0.0000 &  -0.0002 & -0.0004 & -0.0001 & 0.0001 & 0.0001 & -0.0001 \\
          & (0.0027) & (0.0003) & (0.0001) & (0.0006) & (0.0010) & (0.0003) & (0.0006) & (0.0005) & (0.0004) \\
Day$_{-5}$  & -0.0001 & -0.0003 & -0.0001 & -0.0002 & -0.0007 & -0.0003 & 0.0000 & -0.0001 & -0.0002 \\
         & (0.0019) & (0.0004) & (0.0001) & (0.0003) & (0.0012) & (0.0002) & (0.0005) & (0.0004) & (0.0003) \\
Day$_{-4}$  & -0.0005 & -0.0008 & -0.0001 & -0.0007 & -0.0017 & -0.0003 & -0.0007 & 0.0001 & -0.0006 \\
         & (0.0019) & (0.0010) & (0.0001) & (0.0007) & (0.0029) & (0.0002) & (0.0010) & (0.0003) & (0.0007) \\
Day$_{-3}$  & -0.0001 & -0.0003 &  0.0001 & -0.0003 & -0.0012 & 0.0000 & -0.0003 & 0.0001 & -0.0002 \\
         & (0.0019) & (0.0011) & (0.0001) & (0.0007) & (0.0031) & (0.0002) & (0.0009) & (0.0003) & (0.0006) \\
Day$_{-2}$  &  0.0000 &  0.0002 & -0.0001 &  -0.0005 & -0.0004 & -0.0001 & -0.0001 & -0.0002 & 0.0000 \\
         & (0.0019) & (0.0005) & (0.0001) & (0.0003) & (0.0012) & (0.0003) & (0.0006) & (0.0004) & (0.0004) \\
Day$_{-1}$  &  0.0000 &  0.0003 &  0.0000 & -0.0001 &  -0.0003 & 0.0003 & -0.0007 & -0.0006 & 0.0001 \\
         & (0.0023) & (0.0005) & (0.0002) & (0.0008) & (0.0012) & (0.0009) & (0.0017) & (0.0010) & (0.0007) \\
Day$_{0}$  &  0.0010 &  0.0015 &  0.0004 &  -0.0001 &  0.0003 & 0.0007 & 0.0000 & -0.0001 & 0.0009 \\
         & (0.0024) & (0.0008) & (0.0002) & (0.0012) & (0.0015) & (0.0016) & (0.0018) & (0.0013) & (0.0011) \\
Day$_{1}$  &  0.0015 &  0.0023 &  0.0006 &  -0.0004 &  0.0000 & 0.0011 & -0.0016 & -0.0007 & 0.0010 \\
         & (0.0029) & (0.0014) & (0.0005) & (0.0021) & (0.0025) & (0.0023) & (0.0048) & (0.0028) & (0.0024) \\
Day$_{2}$  &  0.0011 &  0.0025 &  0.0006 & -0.0016 &  -0.0003 & 0.0008 & -0.0042 & -0.0011 & 0.0002 \\
          & (0.0038) & (0.0016) & (0.0007) & (0.0028) & (0.0029) & (0.0024) & (0.0076) & (0.0037) & (0.0035) \\
Day$_{3}$  &  0.0004 &  0.0023 &  0.0005 & -0.0025 &  -0.0007 & 0.0002 & -0.0057 & -0.0013 & -0.0007 \\
         & (0.0041) & (0.0017) & (0.0007) & (0.0029) & (0.0029) & (0.0024) & (0.0085) & (0.0041) & (0.0038) \\
Day$_{4}$  &  0.0000 &  0.0022 &  0.0004 & -0.0033 &  -0.0012 & -0.0004 & -0.0063 & -0.0014 & -0.0014 \\
         & (0.0043) & (0.0018) & (0.0008) & (0.0032) & (0.0030) & (0.0025) & (0.0084) & (0.0042) & (0.0040) \\
Day$_{5}$  & -0.0005 &  0.0021 &  0.0004 & -0.0041 & -0.0018 & -0.0012 & -0.0068 & -0.0018 & -0.0022 \\
        & (0.0044) & (0.0019) & (0.0008) & (0.0034) & (0.0031) & (0.0026) & (0.0083) & (0.0043) & (0.0042) \\
Day$_{6}$  & -0.0009 &  0.0020 &  0.0005 & -0.0048 & -0.0021 & -0.0017 & -0.0071 & -0.0021 & -0.0027 \\
        & (0.0043) & (0.0021) & (0.0008) & (0.0036) & (0.0034) & (0.0028) & (0.0087) & (0.0045) & (0.0044) \\
Day$_{7}$  & -0.0010 &  0.0020 &  0.0006 & -0.0051 & -0.0022 & -0.0019 & -0.0071 & -0.0019 & -0.0028 \\
       & (0.0045) & (0.0022) & (0.0008) & (0.0037) & (0.0034) & (0.0029) & (0.0087) & (0.0045) & (0.0045) \\
Day$_{8}$  & -0.0016 &  0.0018 &  0.0005 & -0.0057 & -0.0025 & -0.0022 & -0.0075 & -0.0016 & -0.0033 \\
       & (0.0046) & (0.0022) & (0.0008) & (0.0038) & (0.0035) & (0.0031) & (0.0092) & (0.0046) & (0.0047) \\
Day$_{9}$  & -0.0019 &  0.0017 &  0.0006 & -0.0062 & -0.0026 & -0.0025 & -0.0075 & -0.0014 & -0.0035 \\
       & (0.0046) & (0.0025) & (0.0009) & (0.00039) & (0.0036) & (0.0032) & (0.0091) & (0.0046) & (0.0048) \\
Day$_{10}$  & -0.0018 &  0.0020 &  0.0008 & -0.0064 & -0.0025 & -0.0026 & -0.0070 & -0.0012 & -0.0034 \\
       & (0.0048) & (0.0025) & (0.0009) & (0.0039) & (0.0038) & (0.0034) & (0.0091) & (0.0045) & (0.0048) \\
Day$_{11}$  & -0.0020 &  0.0020 &  0.0008 & -0.0068 & -0.0025 & -0.0028 & -0.0069 & -0.0015 & -0.0033 \\
       & (0.0048) & (0.0026) & (0.0009) & (0.0040) & (0.0039) & (0.0034) & (0.0095) & (0.0045) & (0.0049) \\
Day$_{12}$  & -0.0018 &  0.0023 &  0.0008 & -0.0070 & -0.0023 & -0.0027 & -0.0071 & -0.0018 & -0.0032 \\
       & (0.0049) & (0.0028) & (0.0009) & (0.0041) & (0.0041) & (0.0038) & (0.0097) & (0.0046) & (0.0052) \\
Day$_{13}$  & -0.0020 &  0.0025 &  0.0008 & -0.0076 & -0.0025 & -0.0028 & -0.0080 & -0.0014 & -0.0036 \\
        & (0.0055) & (0.0030) & (0.0010) & (0.0046) & (0.0044) & (0.0042) & (0.0106) & (0.0052) & (0.0056) \\
Day$_{14}$  & -0.0034 & -0.0001 &  0.0005 & -0.0089 & -0.0042 & -0.0063 & -0.0076 & -0.0046 & -0.0049 \\
        & (0.0059) & (0.0035) & (0.0011) & (0.0055) & (0.0047)  & (0.0039) & (0.0118) & (0.0076) & (0.0056) \\
\midrule
Covariates           & {Yes} & {No} & {Yes} & {Yes} & {Yes} & {Yes} & {Yes} & {Yes} & {Yes} \\
Control group        & {Not-yet} & {Not-yet} & {Not-yet} & {Not-yet} & {Never} & {Not-yet} & {Not-yet} & {Not-yet} & {Not-yet} \\
Sub-sample & {No} & {No} & {Top \SI{10}{\percent}} & {Treated} & {No} &{Large} & {Small} & {Pol.}
& {Non-pol.} \\
 &  &  & {removed} & {once} &  &{accounts} & {accounts} & {posts}
& {posts} \\ 

\midrule 
\#Observations       & \num{73836} & \num{73836} & \num{65667} &  \num{39165} & \num{73836} & \num{37170} & \num{36666} & \num{15435} & \num{58401} \\
\#Posts              & \num{3516} & \num{3516} & \num{3127} & \num{1865} & \num{3516} & \num{1770} & \num{1,746} & \num{735} & \num{2781} \\
\bottomrule
\multicolumn{6}{l}{\footnotesize{$^{*}p<0.05$}}
\end{tabular}
\caption{\textbf{Event-study ATTs for Daily Follower Count (Log).} The table displays event-study ATTs for the number of followers (log scale) as outcome variable, from seven days before to fourteen days after receiving a Community Note. Each column presents a different model specification, including sensitivity analyses by follower size and political content. Clustered (bootstrapped) standard errors are reported in parentheses.}
\label{tab:event_time_att_follower_count}
\end{table}

\end{document}